\documentclass[twocolumn,secnumarabic,amssymb, nobibnotes, aps, prd]{revtex4-2}
\usepackage{amsmath}
\usepackage{amsfonts}
\usepackage{amssymb}
\usepackage{graphics}
\usepackage{graphicx}
\usepackage{xcolor}
\providecommand{\U}[1]{\protect\rule{.1in}{.1in}}

\begin{document}
	
\title{Enhancing qubit readout fidelity with two-mode squeezing of the coherent measurement signal}
\author{Baleegh Abdo}
\author{William Shanks}
\author{Oblesh Jinka}
\author{J. R. Rozen}
\affiliation{IBM Quantum, IBM Research Center, Yorktown Heights, New York 10598, USA.}
\date{\today}

\begin{abstract}
The ability to perform high-fidelity quantum nondemolition qubit readout is pivotal for the realization of large and powerful quantum computers. 
Such readout of superconducting qubits is generally enabled by amplifying the weak dispersive measurement signals using phase-preserving quantum-limited Josephson amplifiers with sufficient gain to dilute the contribution of the added noise by the output chain. Here, we further enhance the qubit readout fidelity by (1) simultaneously measuring the two-mode squeezed states of the amplified readout signals at the signal and idler frequencies of the nondegenerate amplifier and (2) coherently combining them at the classical processing stage following a relative rotation that maximizes the signal to noise ratio of the qubit-encoded readout quadrature. 
Such readout scheme exhibits enhancement in the readout fidelity for all practical values of amplifier gain and noise added by the output chain and is fully compatible with frequency multiplexed setups used in large quantum processors.
\end{abstract}

\maketitle
\newpage
\section{Introduction}

Josephson parametric amplifiers (JPAs)  enable fast and high-fidelity Quantum Nondemolition (QND) readout of superconducting qubits, required in large superconducting quantum computers \cite{SCquantumOutlook,BuildingLogicalQubits,VijayQuantumJump,QuantBAScience}. They achieve this by amplifying the weak microwave readout signals carrying the qubit information with a high enough gain to overcome the noise added by subsequent amplification stages in the output chains, while adding near the minimum amount of noise required by quantum mechanics to the processed input signals \cite{Caves,NoiseReview}. 

Thus, to build a frequency-multiplexed output chain suitable for high-fidelity dispersive readout \cite{cQED1,cQED2}, two main components are needed in the ideal case, (1) a  directional JPA at $10$ mK, that has a high power gain ($>20$ dB), adds only half a photon of input noise, and lacks classical backaction, followed by (2) a high-gain ($>40$ dB), low-noise amplifier (e.g., HEMT) at $4$ K that adds as few input noise photons as possible \cite{ScPAs}. 

In practice, however, most output chains exhibit much worse gain and noise performance. This is primarily due to unwanted insertion loss in the output chain and non-ideal performance of the wideband JPA used, which typically takes the form of a Traveling Wave Parametric Amplifier (TWPA) \cite{TWPAScience,FloquetTWPA,LowLossTWPA}. Typically, losses in the output chain originate from dissipation in the JPA (e.g., TWPA \cite{LowLossTWPA}) and the intermediate microwave components, such as magnetic isolators, attenuators, and filters, which are added to protect the qubits against thermal noise coming from higher temperature stages and  classical backaction of the two amplifiers, as well as to dampen reflections between unmatched components. Consequently, such insertion loss, which can be in the $3-10$ dB range, considerably elevates the effective added noise photons at the input of the HEMT to $20-100$. Similarly, non-idealities of common TWPAs \cite{TWPAScience} which originate from undesired high-order mixing products, impedance mismatch with other components, and gain instabilities \cite{FloquetTWPA2}, further degrade the performance of output chains, for example by requiring a TWPA gain below $20$ dB to maintain a stable continuous operation. 

In recent years several measurement schemes were proposed or tested, which can potentially  boost the readout fidelity of superconducting qubits. Examples of such schemes include using interferometric squeezing schemes for the readout signals formed by two degenerate or nondegenerate JPAs that precede and succeed the quantum processor \cite{InterDisMeas,JPCsqueezingReadout,StrobQuMeas}, applying external and intercavity single-mode squeezing \cite{SqInAndOut}, devising longitudinal interactions between the qubits and readout resonators \cite{LongitudinalRead,LongReadYale,LongReadFin,EnabledQubitCloaking}, coupling the qubit to two readout resonators that are measured with two-mode squeezed states generated internally \cite{IntTwoModeSq} or externally \cite{ExtTwoModeSq}, coupling the qubit to a JPA on chip \cite{ReadoutAmpl}, using a coupled readout resonator-Purcell filter system and fast flux tuning of the qubit frequency \cite{WallraffFastRead}, exploiting higher energy levels of the qubit \cite{ReadoutWoJPA}, employing cross-Kerr nonlinear effects enabled by additional nonlinear modes or elements  \cite{MoleculeCrossKerr,QuatronCoupler,EfficientDecoupling,CrossKerrJJreadout}, and applying diversity combining techniques that typically enhance the signal-to noise ratio (SNR) of noisy classical signals \cite{MicrosoftPaper,LinearDiversityComb}.     

While most of these schemes are viable and promising, they generally come with serious drawbacks such as increased complexity and control, added hardware overhead, and enhanced off-chip losses. Moreover, some of them require making major changes to the quantum processor architecture itself.

In this work, we demonstrate a much simpler enhanced readout method that largely uses the same resources of state-of-the-art readout output chains. It relies on three important properties of qubit readout measured with phase-preserving, near quantum limited amplifiers (QLAs), such as TWPAs, JPAs \cite{JPASQUIDArrayJILA,JPAHat,JPAimpedanceEng,BroadbandCPWJPA}, and nondegnerate Josephson mixers (JMs) \cite{JPCreview,hybridLessJPC,Roch,microstripJPC,JPCnature}, namely (1) the information about the qubit state is primarily encoded in one quadrature of the coherent readout signal, (2) the noise added by the output chain following the QLA is much larger than the quantum limit, and (3) phase-preserving amplification in which both quadratures of the processed signal are equally amplified can yield squeezing and antisqueezing of certain combinations of quadratures of the entangled output modes (two-mode squeezed states) generated by the amplifier at the signal ($f_s$) and idler ($f_i$) frequencies. 

Using these common properties, we obtain an enhancement in the SNR of our measurement chain and in turn the readout fidelity by coherently combining simultaneous qubit measurements at $f_s$ and $f_i$ (referred back to the output of the QLA) such that the qubit information is contained in the antisqueezed combined quadrature. While generally the idler mode of the amplifier does not contain additional information about the qubit (particularly in the high-gain limit), the antisqueezed combined quadrature, which equally enhances the signal and noise of the QLA, tends to increase the total SNR in the presence of additional noise in the output chains. 
             
Although the proposed measurement scheme resembles the operation of first-stage, phase-sensitive QLAs, the latter does not require the addition of input noise originating from the idler port to the antisqueezed quadrature. Our method, on the other hand, does not suffer from two limitations of phase-sensitive amplifiers that render them unsuitable for frequency-multiplexed readout schemes. In particular, our method does not require any in-situ phase matching between the pump and the amplified quadrature of the measurement signal, and most importantly, it does not apply to a single operational frequency set by the pump frequency $f_p$ and the degeneracy of the signal and idler frequencies, which satisfy $f_s=f_i=f_p$ ($f_s=f_i=f_p/2$) in the case of a four-wave (three-wave) mixing nonlinearity. Furthermore, in contrast to the last limitation, our method can be simultaneously applied to any nondegenerate signal and idler frequencies (i.e., $f_s \neq f_i$) within the device bandwidth, which satisfy $2f_p=f_s+f_i$ ($f_p=f_s+f_i$) when using a four-wave (three-wave) mixing nonlinearity. 

\section{Results}
\subsection{Readout scheme}

\begin{figure*}
	[tb]
	\begin{center}
		\includegraphics[
		width=2\columnwidth 
		]%
		{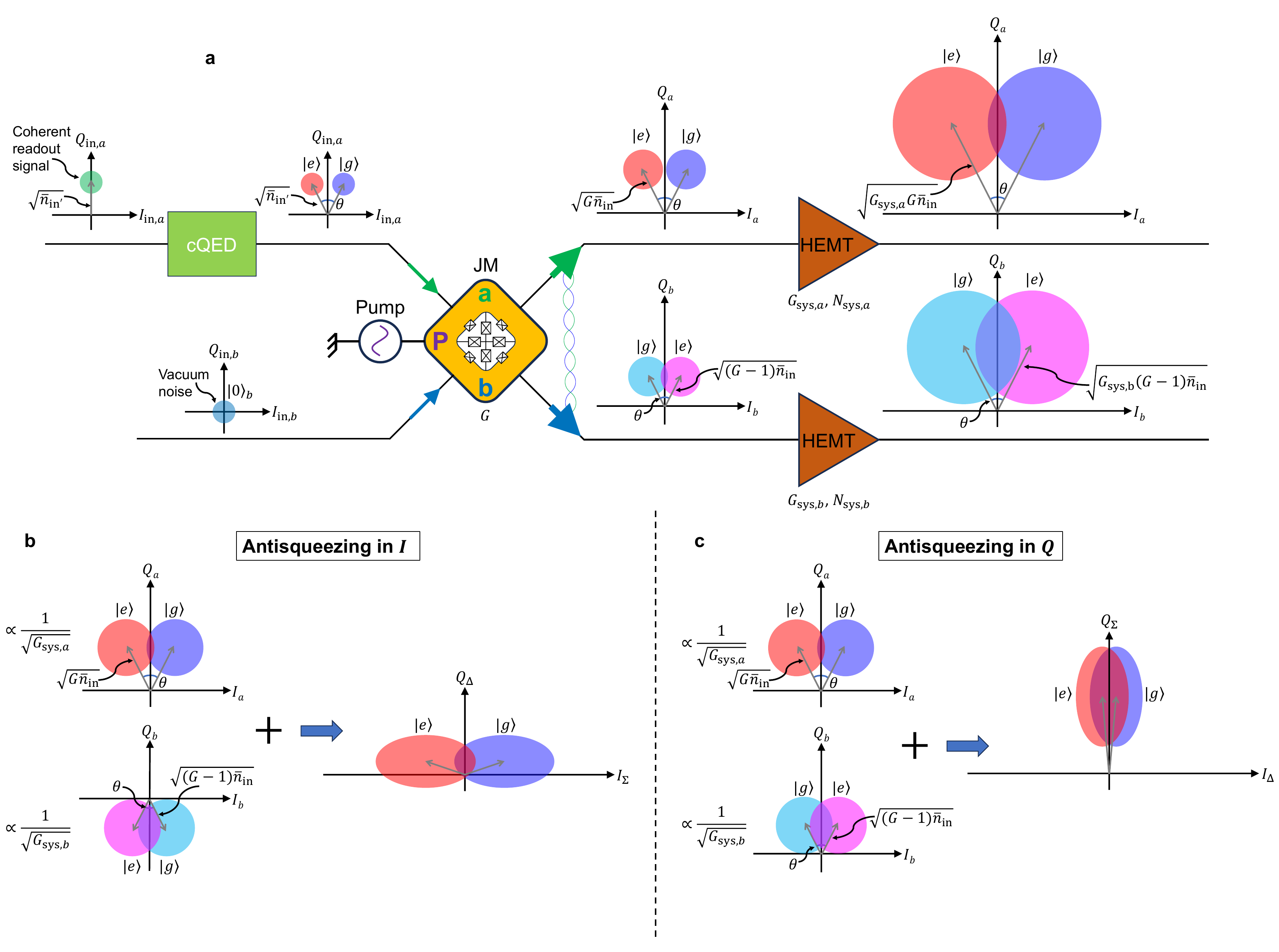}
		\caption{\textbf{Harnessing two-mode squeezing of amplified qubit readout signals.} \textbf{a} Illustration of a dispersive qubit readout signal amplified by a nondegenerate quantum-limited amplifier (e.g., JM) connected to two output lines. Antisqueeezing in the $I$ ($Q$) quadrature carrying (lacking) the qubit state information can be obtained in scenario \textbf{b} (\textbf{c}) by coherently combining the simultaneous $IQ$ measurements referred back to the output ports $a$ and $b$ of the JM and rotated by a relative phase of $\pi$ ($0$). Using the processing protocol of \textbf{b} (\textbf{c}) we get an enhancement (reduction) in the qubit readout fidelity in comparison to the standard readout with phase-preserving quantum-limited amplification (represented by the measurement result depicted in the upper arm of \textbf{a}). In this illustration, we use low $G=2$, $G_{\rm{sys}}=4$, and $N_{\rm{sys}}=1$ to keep the size of the 2D scatter at the output of the HEMTs relatively small so that the main features of the readout method are not obscured. We also use $\theta=54^{\circ}$. 		    
		}
		\label{TwoModeSqReadout}
	\end{center}
\end{figure*}

To demonstrate the proposed readout scheme we use a nondegenerate, three-wave JM coupled to the output of a circuit quantum electrodynamics (cQED) device, consisting of a transmon qubit capacitively coupled to a superconducting readout resonator, as illustrated in Fig.\,\ref{TwoModeSqReadout} \textbf{a}. The JM operation is based on parametric modulation of a dispersive nonlinear three-wave mixing medium, known as the Josephson ring modulator (JRM), which couples to two nondegenerate microwave resonators denoted $a$ and $b$ that have two separate ports. One common realization of the JRM,  depicted in the JM symbol in Fig.\,\ref{TwoModeSqReadout}, consists of four Josephson junctions arranged in a Wheatstone bridge configuration and shunted with linear inductors in the form of four large JJs inside the loop \cite{hybridLessJPC}. 

Like standard dispersive readout schemes, we apply a pulsed single-tone, coherent microwave signal to the readout resonator of the cQED, represented as a displaced disk along the $Q$ axis in the $IQ$ plane. In this picture, the distance between the disk center and the origin corresponds to $\sqrt{\bar{n}_{\rm{in'}}}$, where $\bar{n}_{\rm{in'}}$ is the average photon number populating the readout resonator, and the disk radius represents the standard deviation of vacuum noise, which equals $1/2$ in quanta units. Upon interaction with qubit, the readout coherent state undergoes a qubit-state dependent phase rotation, $\pm \theta/2$, as shown at the cQED output, which distinguishes between the ground and excited states of the qubit.

In an analogous manner to other widely used phase-preserving QLAs, the JM equally amplifies both quadratures of the readout signals input on port $a$ by an amplitude gain $\sqrt{G}$, while adding, in the high-gain limit, half a photon of noise, originating from the idler vacuum noise on port $b$. The average input photon number reaching the JM input $\bar{n}_{\rm{in}}=\bar{\alpha}\bar{n}_{\rm{in'}}$, depends on  the insertion loss $0<\bar{\alpha}\leq 1$ between the cQED and the JM. In standard high-fidelity qubit readout schemes, represented by the top path in Fig.\,\ref{TwoModeSqReadout} \textbf{a}, this phase-preserved amplified readout signal is amplified further by a low-noise amplifier, i.e., the HEMT at $4$ K, and subsequently is downconverted, sampled, and digitized using room-temperature electronics. For simplicity, we attribute the total gain of the output chain (in the absence of the JM) to the HEMT, which we denote $G_{\rm{sys},a}$. We also refer the added noise by the output chain back to the HEMT input, which we denote $N_{\rm{sys},a}$.   

Unlike standard dispersive readout schemes, however, we also measure the amplified readout signal on port $b$ of the JM, represented by the bottom path in Fig.\,\ref{TwoModeSqReadout} \textbf{a}. As illustrated in Fig.\,\ref{TwoModeSqReadout} \textbf{b} and \textbf{c}, we exploit the entangled modes at the output of the JM (i.e., the two-mode squeezed state) \cite{GenEntMwRadoverTL,QuantumNode,EntSwap} to either amplify the information encoded in one quadrature $I$ and squeeze the other $Q$ (Fig.\,\ref{TwoModeSqReadout} \textbf{b}) or vice versa (Fig.\,\ref{TwoModeSqReadout} \textbf{c}). The subscripts $\Sigma$ and $\Delta$ are employed to indicate antisqueezing and squeezing of the combined $IQ$ quadratures, respectively (see Supplementary Information). 
  
As shown in Figs.\,\ref{TwoModeSqReadout} \textbf{b} and \textbf{c}, the processing protocol consists of four main steps: (1) rotating the integrated $I$ and $Q$ data points taken using the two output lines by a fixed phase, which causes the centroids of the 2D scatter of the $\left| g\right\rangle $ and $\left| e\right\rangle $ states to lie symmetrically around the $Q$ axis (as shown on the right side of Figs.\,\ref{TwoModeSqReadout} \textbf{a}), (2) scaling the corresponding $I$ and $Q$ data points back to the output of ports $a$ and $b$ of the JM. This is done by converting the measured results into the photon basis (see Methods and Sec. IV in the Supplementary Information), (3) applying a relative rotation of $\pi$ ($0$) to the data points measured on output line $b$ (obtained using the trans-gain process in the JM) as shown in Fig.\,\ref{TwoModeSqReadout} \textbf{b} (Fig.\,\ref{TwoModeSqReadout} \textbf{c}), and (4) coherently combining the two data sets, represented by the coordinates ($I_a$, $Q_a$) and ($I_b$, $Q_b$). 

When comparing the resultant 2D scatter on the right side of Fig.\,\ref{TwoModeSqReadout} \textbf{b} (Fig.\,\ref{TwoModeSqReadout} \textbf{c}), to that of the standard phase-preserving amplification shown, for example, on the top left side of Fig.\,\ref{TwoModeSqReadout} \textbf{b}, we generally find that the inferred readout fidelity, which we quantify below, increases (decreases) in the case of antisqueezing of $I$ (antisqueezing of $Q$). We also find that the distance of the centroids relative to the origin  generally decreases (increases) in the case of antisqueezing of $I$ (antisqueezing of $Q$) compared to the  standard phase-preserving amplification case.

\begin{figure*}
	[tb]
	\begin{center}
		\includegraphics[
		width=1.62\columnwidth 
		]%
		{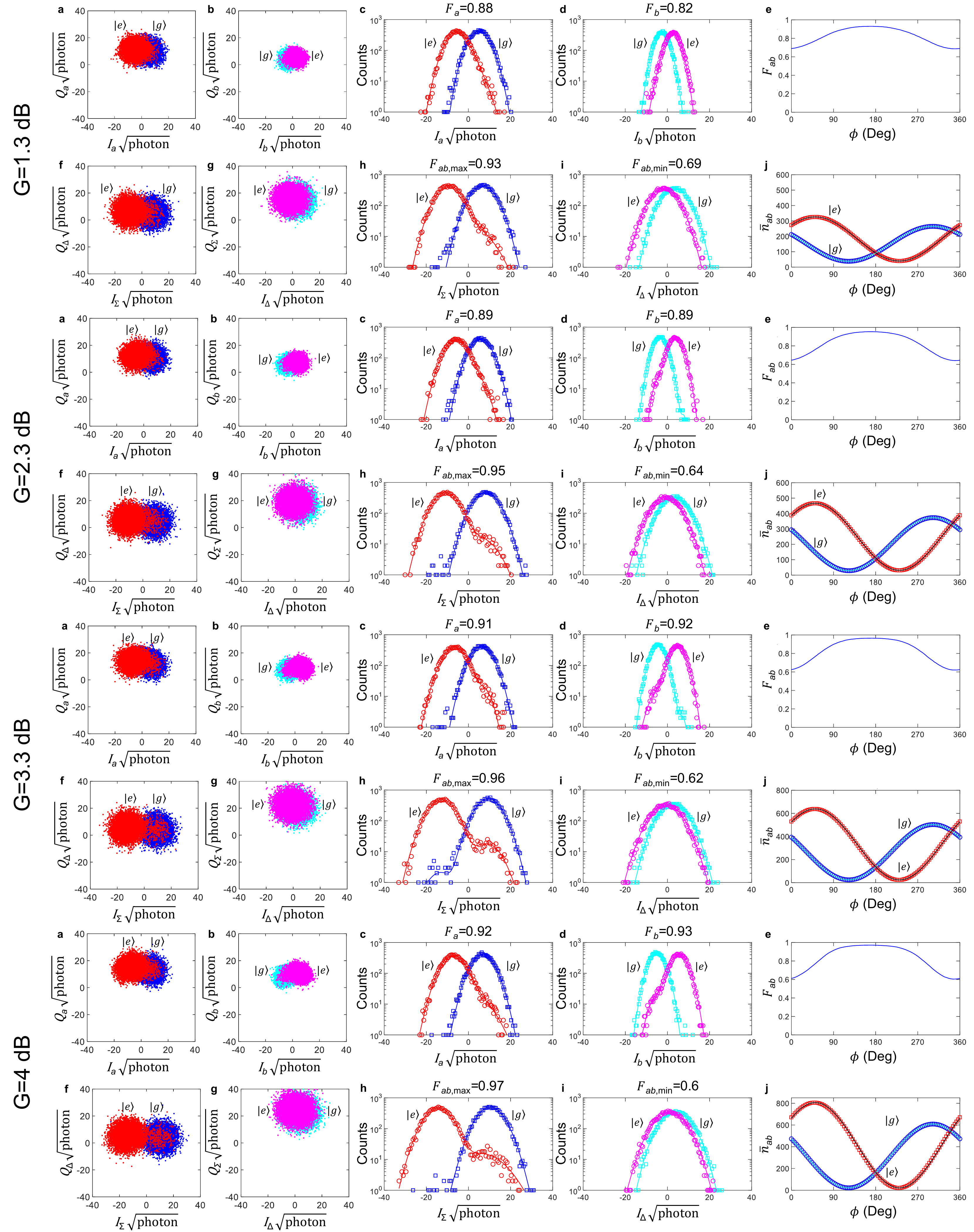}
		\caption{\textbf{Qubit readout fidelity measured using mode \textit{a}, \textit{b} and the combined mode \textit{ab} of the JM for different gains.} The following applies to the various gains listed on the left side of the figure. \textbf{a} (\textbf{b}) Readout measurement shots in the $IQ$ plane taken with mode \textit{a} (\textit{b}) corresponding to initialization of the qubit in the $g$ and $e$ states. \textbf{c} (\textbf{d}) Histograms of the data in \textbf{a} (\textbf{b}) taken along quadrature $I$. \textbf{e} The readout fidelity of the combined mode \textit{ab} versus the relative rotation angle $\phi$. \textbf{f} (\textbf{g}) same as \textbf{a} (\textbf{b}) obtained for the combined mode \textit{ab} with $\phi=\pi$ ($\phi=0$). \textbf{h} (\textbf{i}) same as \textbf{c} (\textbf{d}) corresponding to the data in \textbf{f} (\textbf{g}). \textbf{g} The extracted average photon number of the combined mode for states $g$ and $e$ versus $\phi$. The solid curves in the various histogram plots correspond to double Gaussian fits. All measurements are taken at a fixed $\bar{n}_{\rm{in}}=90$. 
		}
		\label{HistandFidpanels}
	\end{center}
\end{figure*}

\subsection{Experimental results}

\begin{figure*}
	[tb]
	\begin{center}
		\includegraphics[
		width=2\columnwidth 
		]%
		{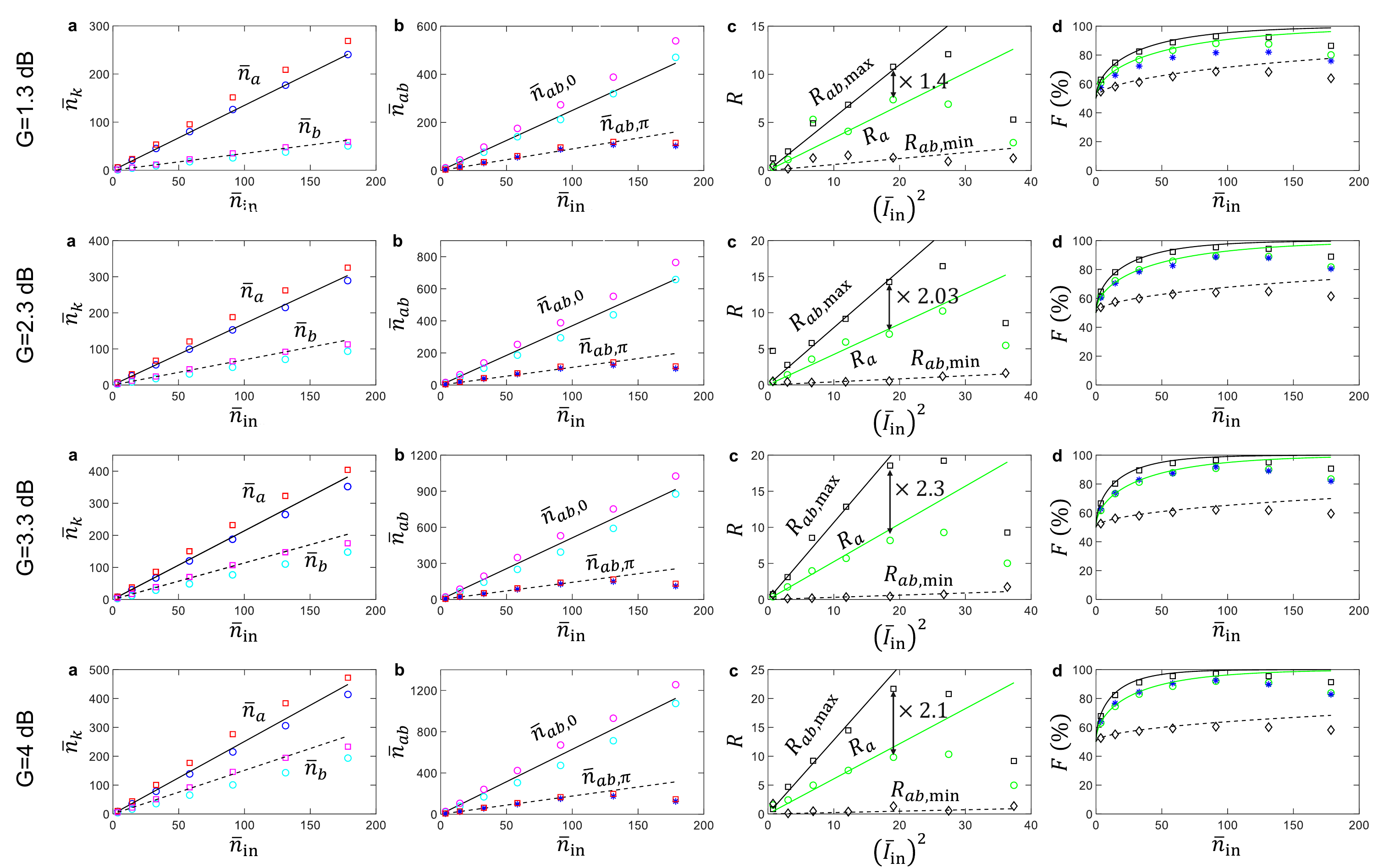}
		\caption{\textbf{The average photon number at the JM output, $R$, and $F$ versus $\bar{n}_{\rm{in}}$.} The following applies to the various gains listed on the left side of the figure. \textbf{a} The data represent measured average photon numbers at the JM output for mode $a$ ($\bar{n}_a$) and $b$ ($\bar{n}_b$) corresponding to the qubit $g$ (blue and cyan) and $e$ states (red and magenta). The solid and dashed black lines are fits based on Eqs.\,(\ref{n_a_bar_text}) and (\ref{n_b_bar_text}). \textbf{b} is similar to \textbf{a}. The data represent the fictitious average photon number at the JM output for the combined mode $ab$ corresponding to $\phi=0$ ($\bar{n}_{ab,0}$) and $\phi=\pi$ ($\bar{n}_{ab,\pi}$). The dashed and solid black lines are fits based on Eqs.\,(\ref{n_ab_bar_pi_r_text}) and (\ref{n_ab_bar_zero_r_text}). \textbf{c} The power SNR ($R$) versus $\bar{I}^2_{\rm{in}}$. The circles, squares, and diamonds correspond to $R$ measured for mode $a$, the combined mode $ab$ that maximizes $R$, and minimizes it, respectively. The green and black solid lines and the black dashed line are theory fits based on Eqs.\,(\ref{R_a_uniform_r_text}), (\ref{R_ab_max_text}), and (\ref{R_ab_min_text}), respectively. \textbf{d} Assignment fidelity versus $\bar{n}_{\rm{in}}$ measured with mode $a$ (green circles), mode $b$ (blue stars), combined mode $ab$ that maximizes $R$ (black squares), and minimizes it (black diamonds). The green and black solid curves and the black dashed curve are theory fits obtained by substituting the results of  Eqs.\,(\ref{R_a_uniform_r_text}), (\ref{R_ab_max_text}), and (\ref{R_ab_min_text}) in Eq.\,(\ref{F_vs_R}), respectively. 
		}
		\label{NbarRandFvsNin}
	\end{center}
\end{figure*}

In Fig.\,\ref{HistandFidpanels}, we exhibit the measured 2D scatter and assignment fidelities corresponding to four JM gains $G=1.3$, $2.3$, $3.3$, $4$ dB taken at a fixed $\bar{n}_{\rm{in}}=90$. For each gain working point, we display in panels \textbf{a} and \textbf{b} the 2D scatter for the ground and excited states measured at outputs $a$ (reflection gain) and $b$ (transmission gain) of the JM. In panels \textbf{c} and \textbf{d}, we display the measured histograms taken along the $I$ quadrature and the corresponding double Gaussian fits (solid curves). We also list at the top of the panels the corresponding assignment fidelities $F_{a}$ and $F_{b}$ obtained from the data, where $F=1-\left[P\left(g |e \right)+P\left(e |g \right)  \right] /2$. In panel \textbf{e}, we plot $F_{ab}$ the calculated assignment fidelity of the combined data sets of panels \textbf{a} and \textbf{b} as a function of the relative phase $\phi$ applied to the trans-gain data in panels \textbf{b}.
 
Similarly, in panels \textbf{f} and \textbf{g}, we plot the resultant 2D scatter of the combined data sets of panels \textbf{a} and \textbf{b} obtained at $\pi$ and $0$ relative phase rotations, respectively, which correspond to the cases of antisqueezing of $I$ and $Q$ illustrated in Fig.\,\ref{TwoModeSqReadout} \textbf{b} and Fig.\,\ref{TwoModeSqReadout} \textbf{c}. In panels \textbf{h} and \textbf{i}, we depict the histograms of the combined data sets of panels \textbf{f} and \textbf{g} obtained along the $I$ quadrature and the double Gaussian fits (solid curves). We also list at the top of the panels the corresponding assignment fidelities $F_{ab,\rm{max}}$ and $F_{ab,\rm{min}}$ obtained from the data. 

Lastly, in panels \textbf{j}, we plot the fictitious average photon number of the combined mode versus $\phi$ (see Sec. II in the Supplementary Information)

\begin{align}
	\bar{n}_{ab,i}\left(\phi \right) =\bar{I}^2_{ab,i}\left(\phi \right)+\bar{Q}^2_{ab,i}\left(\phi \right), \label{n_ab_i_bar_text} 
\end{align}

\noindent corresponding to the ground $i=g$ (blue) and excited $i=e$ (red) states along with their sinusoidal fits (solid curves).  

From these results, we observe that (1) $F_{a}\cong F_{b}$ as expected when reading out the qubit using two entangled outputs (as seen in panels \textbf{c} and \textbf{d}), (2) the various assignment fidelities generally
satisfy $F_{ab,\rm{min}} < F_{a}\cong F_{b}< F_{ab,\rm{max}}$ (as seen in panels \textbf{c}, \textbf{d}, \textbf{h}, \textbf{i}), (3) the combined assignment fidelity $F_{ab}$ reaches a relatively broad maximum (minimum) around $\phi=180$ ($\phi=0$) degrees (as seen in panels \textbf{e}), and (4) $\bar{n}_{ab}$ is simultaneously minimized (maximized) for the $\left| g \right\rangle $ and $\left| e \right\rangle $ states around $\phi=180$ ($\phi=0$) degrees (as seen in panels \textbf{j}). 

\begin{figure*}
	[tb]
	\begin{center}
		\includegraphics[
		width=1.9\columnwidth 
		]%
		{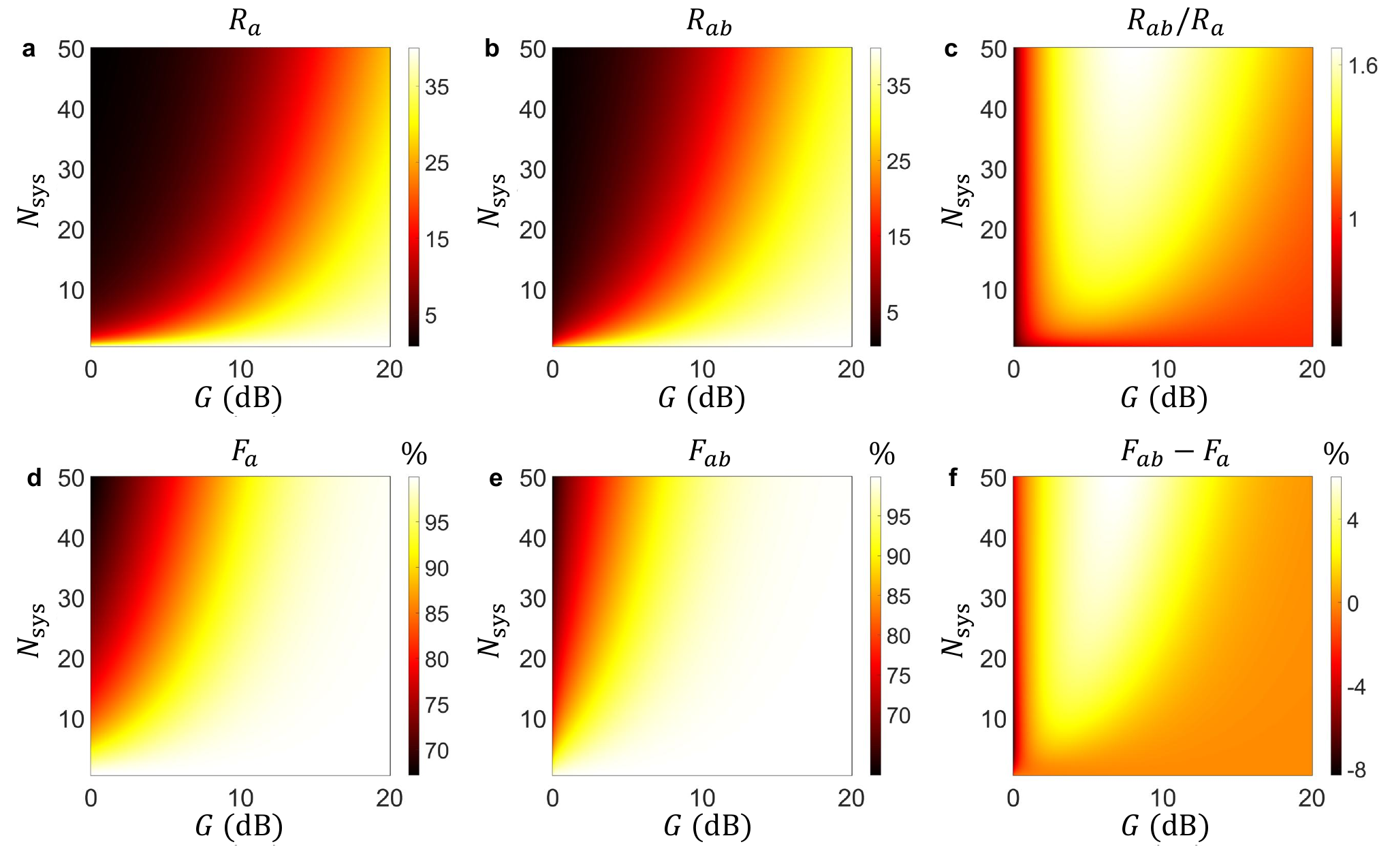}
		\caption{\textbf{Calculated dependence of $R$ and $F$ on JM gain $G$ and added noise by the output chains $N_{\rm{sys}}$.} \textbf{a} $R$ obtained for mode $a$  (Eq.\,(\ref{R_a_uniform_r_text})). \textbf{b} maximum $R$ obtained for the combined mode $ab$ (Eq.\,(\ref{R_ab_max2_text})). \textbf{c} Calculated ratio $R_{ab}/R_a$ of the metrics in panels \textbf{b} and \textbf{a}. \textbf{d} $F$ obtained for mode $a$  (using Eqs.\,(\ref{R_a_uniform_r_text}) and (\ref{F_vs_R})). \textbf{e} maximum $F$ obtained for the combined mode $ab$ (using Eqs.\,(\ref{R_ab_max2_text}) and (\ref{F_vs_R})). \textbf{c} Calculated difference, i.e., $F_{ab}-F_a$, of the fidelities in panels \textbf{e} and \textbf{d}. In panels \textbf{a}, \textbf{b}, \textbf{d}, \textbf{e}, $\bar{I}^2_{\rm{in}}=5$ is used. All calculations are done for equal added noise by the output chains, i.e., $N_{\rm{sys}} \equiv N_{\rm{sys},a}=N_{\rm{sys},b}$.        
		}
		\label{RandFvsGandN}
	\end{center}
\end{figure*}

\begin{figure}
	[tb]
	\begin{center}
		\includegraphics[
		width=\columnwidth 
		]%
		{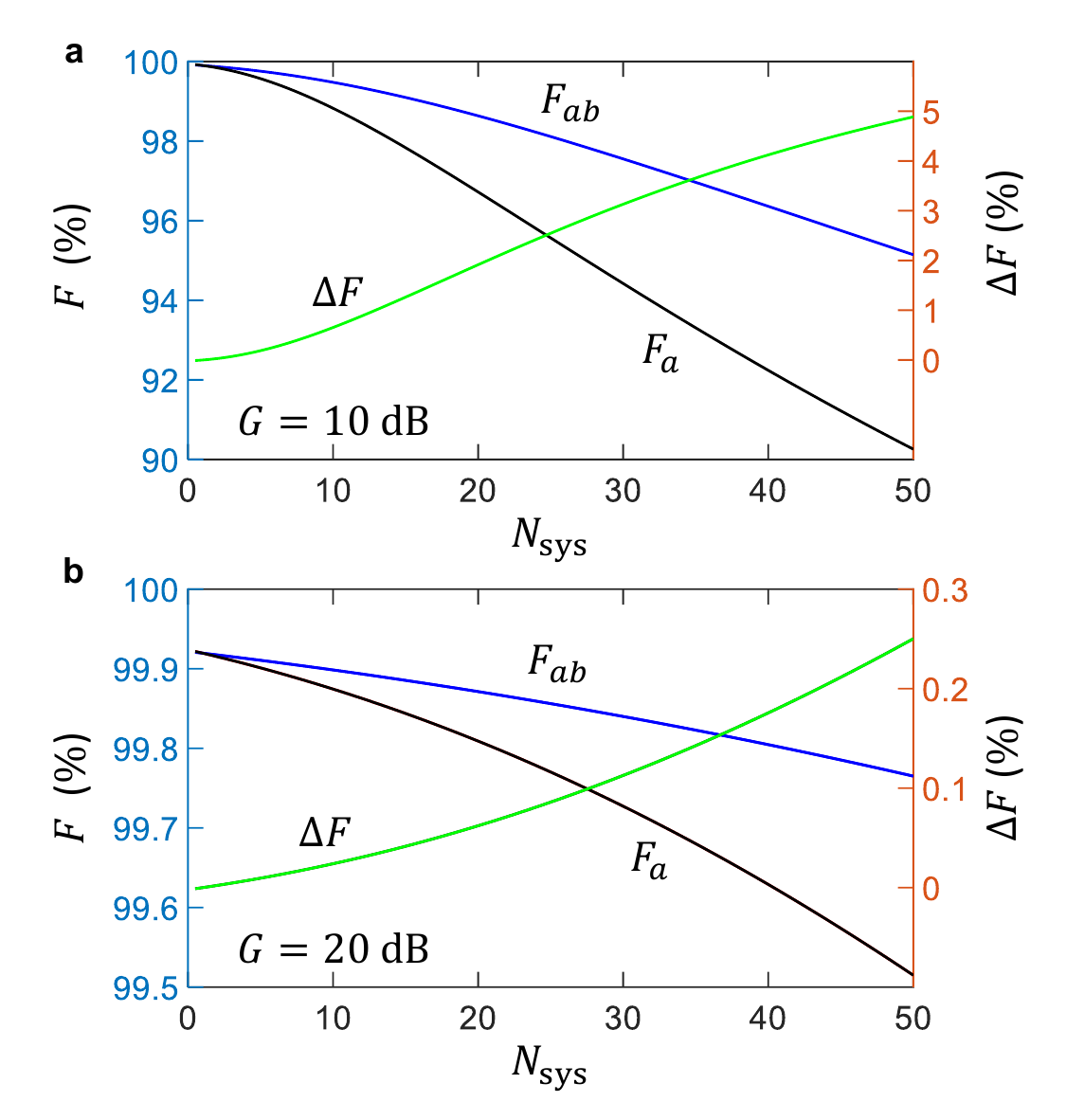}
		\caption{\textbf{Calculated fidelity versus $N_{\rm{sys}}$ for fixed gains.} \textbf{a} and \textbf{b} show cross sections of the calculated $F_a$ (black), $F_{ab}$ (blue), and $\Delta F=F_{ab}-F_{a}$ versus $N_{\rm{sys}}$ of Fig.\,\ref{RandFvsGandN}, taken at fixed JM gains of $10$ dB (\textbf{a}) and $20$ dB (\textbf{b}).     
		}
		\label{Fcrosssections}
	\end{center}
\end{figure}

It is important to note that in the experiment, the JM gain is limited to low levels due to the finite isolation of about $18$ dB of a single circulator between the cQED and the JM (see the experimental setup depicted in Fig. S2). As shown in Fig. S3 considerable backaction effects such as diminished coherence times of the qubit, i.e., $T_1$, $T_{2E}$, are observed at higher gains.

In Fig.\,\ref{NbarRandFvsNin}, we plot for the JM gains $G=1.3$, $2.3$, $3.3$, $4$ dB (displayed in different rows) various important metrics that we calculate from the measured 2D scatter shown in Fig.\,\ref{HistandFidpanels}. In panels \textbf{a}, we plot the average photon number at the output of the JM at port $k=a,b$ versus $\bar{n}_{\rm{in}}$, where the data points are calculated for each qubit state $i=g,e$ using the relation (see Supplementary information)

\begin{align}
	\bar{n}_{k,i}=\bar{I}^2_{k,i}+\bar{Q}^2_{k,i}, \label{n_m_i_bar} 
\end{align}

\noindent while the theory fits drawn as solid and dashed black lines are obtained using the standard phase-preserving amplification relations 

\begin{align}
	\bar{n}_{a}=\cosh^{2}(r)\bar{n}_{\rm{in}}, \label{n_a_bar_text} \\
	\bar{n}_{b}=\sinh^{2}(r)\bar{n}_{\rm{in}}, \label{n_b_bar_text}
\end{align}

\noindent where $G=\cosh^2\left( r\right)$ with the squeezing parameter $r\geq 0$. 

Likewise, in panels \textbf{b}, we plot the average photon number of the combined mode $\bar{n}_{ab}$ versus $\bar{n}_{\rm{in}}$ for $\phi=0$ and $\phi=\pi$ calculated using Eq.\,(\ref{n_ab_i_bar_text}). Whereas the theory fits drawn using solid and dashed lines are calculated using (see Supplementary information)

\begin{align}
	\bar{n}_{ab,0}=e^{-2r} \bar{I}^2_{\rm{in}}+e^{2r} \bar{Q}^2_{\rm{in}}, \label{n_ab_bar_zero_r_text} 
\end{align}

\noindent and 

\begin{align}
	\bar{n}_{ab,\pi}=e^{2r} \bar{I}^2_{\rm{in}}+e^{-2r} \bar{Q}^2_{\rm{in}}, \label{n_ab_bar_pi_r_text} 
\end{align}
 
\noindent respectively, where $\bar{I}_{\rm{in}}=\sin\left( \theta/2\right)\sqrt{\bar{n}_{\rm{in}}}$ and $\bar{Q}_{\rm{in}}=\cos\left( \theta/2\right)\sqrt{\bar{n}_{\rm{in}}}$ are the quadratures of the input signal at port $a$. As seen in Eq.\,(\ref{n_ab_bar_zero_r_text}) (and vice versa in Eq.\,(\ref{n_ab_bar_pi_r_text})),  $\bar{n}_{ab,0}$ is obtained by squeezing of $\bar{I}_{\rm{in}}$ by $e^{-r}$ and antisqueezing of $\bar{Q}_{\rm{in}}$ by $e^{r}$.  

Note that the results in Fig.\,\ref{NbarRandFvsNin} \textbf{b} are in agreement with the observed behavior of $\bar{n}_{ab}$ versus $\phi$ shown in Fig.\,\ref{HistandFidpanels} \textbf{j}, in which $\bar{n}_{ab}$ is jointly minimized (maximized) for $\left| g\right\rangle $ and $\left| e\right\rangle $ at $\phi=\pi$ ($\phi=0$).

Next, we compare the performance of the different measurement methods with respect to the power SNR denoted $R$ and assignment fidelity versus the input power taken at different JM gains.

The power SNR of the qubit-state measurement is given by

\begin{equation}
	R\equiv\rm{SNR}^2=\left( \dfrac{\bar{I}_g-\bar{I}_e}{\dfrac{\sigma_g+\sigma_e}{2}}\right)^2, \label{R_def_text}
\end{equation}

\noindent where $\bar{I}$ ($\sigma$) is the mean (standard deviation) of the $I$ quadrature of the output coherent state carrying the qubit information. 

Assuming for simplicity that $\bar{I}_g=-\bar{I}_e$ and $\sigma\equiv\sigma_g=\sigma_e$, we get

\begin{equation}
	R=\left( \dfrac{2\bar{I}}{\sigma}\right)^2, \label{R_uniform_text}
\end{equation}

\noindent where $\left| \bar{I}\right| \equiv\left|\bar{I}_g\right|=\left|\bar{I}_e\right|$.

Using Eq.\,(\ref{R_uniform_text}), we obtain for the measurement with mode $a$ (see Supplementary information) 

\begin{equation}
	R_a=\dfrac{\left( e^{2r}+e^{-2r}+2 \right) }{\left( e^{2r}+e^{-2r}\right) /8+N_{\rm{sys},a}/2}\bar{I}^2_{\rm{in}}. \label{R_a_uniform_r_text}
\end{equation}

Similarly, we obtain for the measurement with  the combined mode $ab$ in the case of antisqueezed $I$ quadrature (see Supplementary information) 

\begin{equation}
	R_{ab,\rm{max}}=\dfrac{2e^{2r} }{e^{2r}/4+N_{\rm{sys},e}/4}\bar{I}^2_{\rm{in}}, \label{R_ab_max_text}
\end{equation}

\noindent and  

\begin{equation}
	R_{ab,\rm{min}}=\dfrac{2e^{-2r} }{e^{-2r}/4+N_{\rm{sys},e}/4}\bar{I}^2_{\rm{in}}, \label{R_ab_min_text}
\end{equation}

\noindent in the case of antisqueezed $Q$ quadrature, where $N_{\rm{sys},e}$ is an effective added noise photons by the output chains for modes $a$ and $b$. Assuming uncorrelated Gaussian noise with zero mean for the two output chains, we obtain $N_{\rm{sys},e}=N_{\rm{sys},a}+N_{\rm{sys},b}$. In the special case of equal added noise by the two output chains, i.e., $N_{\rm{sys},a}=N_{\rm{sys},b}$, Eq.\,(\ref{R_ab_max_text}) simplifies to 

\begin{equation}
	R_{ab,\rm{max}}=\dfrac{2e^{2r} }{e^{2r}/4+N_{\rm{sys},a}/2}\bar{I}^2_{\rm{in}}. \label{R_ab_max2_text}
\end{equation}
 
When comparing Eq.\,(\ref{R_a_uniform_r_text}) and Eq.\,(\ref{R_ab_max2_text}), we find that: (1) for $r\gg 0$, both the signal and noise powers amplified by the JM are two times higher in the combined mode measurement than in the phase-preserving amplification case while the added noise contribution in the denominator is equal, (2) for unity gain ($r=0$), $R_a$ is twice $R_{ab}$ because with no trans-gain process, the combined mode measurement combines noise of two modes while the signal is present in only one, (3) in the absence of any added noise by the output chains (i.e., $N_{\rm{sys,a}}=0$) or in the infinite gain limit, $R_a$ and $R_{ab,\rm{max}}$ are equal as expected, and (4) $R_{ab,\rm{max}}$ exceeds $R_{a}$ in most practical cases in which $N_{\rm{sys,a}}\gg 1/2$ and $r\gg0.35$ ($G\gg0.5$ dB). 

In panels \textbf{c} of Fig.\,\ref{NbarRandFvsNin}, we plot $R$ given by Eq.\,(\ref{R_def_text}) as a function of $\bar{I}^2_{\rm{in}}$, which we calculate using the histograms of mode $a$ (green circles), the combined mode $ab$ for $\phi=\pi$ (black squares) and $\phi=0$ (black diamonds). As seen in the latter two cases, which correspond to the antisqueezing of $I$ and antisqueezing of $Q$, $R$ is maximized and minimized, respectively. Enhancements of more than a factor of two are observed in $R_{ab,\rm{max}}$ versus $R_{a}$ at around $\bar{I}^2_{\rm{in}}=20$ for $G=2.3, 3.3, 4$ dB. Such enhancements in the SNR of the combined mode $ab$ relative to mode $a$ are partly enabled by the lower added noise of  output chain $b$ compared to $a$ in our experiment, i.e., $N_{\rm{sys,a}}=28.2$ versus $N_{\rm{sys,b}}=13.4$ (See Table I in the Supplementary Information). The respective theory fits are drawn for $R_a$ (green solid line), $R_{ab,\rm{max}}$ (black solid line), and $R_{ab,\rm{min}}$ (black dashed line) using Eq.\,(\ref{R_a_uniform_r_text}), (\ref{R_ab_max_text}), (\ref{R_ab_min_text}), respectively. The observed deviation of the measured data points from the theory fits at elevated input powers $\bar{I}^2_{\rm{in}}>20$ can be due to increased adverse effects taking place in the cQED, such as relaxation or backaction.  

In panels \textbf{d} of Fig.\,\ref{NbarRandFvsNin}, we calculate the assignment fidelity for modes $a$ (green circles), $b$ (blue stars) and $ab$ for $\phi=\pi$ (black squares) and $\phi=0$ (black diamonds) versus $\bar{n}_{\rm{in}}$. The various fidelity data points are extracted from measured histograms similar to the ones depicted in Fig.\,\ref{HistandFidpanels}. Whereas, the theory fits are calculated using the known relation

\begin{equation}
	F=\dfrac{1}{2}\left[1+\rm{erf}\left(\sqrt{\dfrac{R}{8}}\right)  \right], \label{F_vs_R}
\end{equation}

\noindent and the theoretical values of $R$ obtained using Eq.\,(\ref{R_a_uniform_r_text}) for $R_a$ (green solid curve), Eq.\,(\ref{R_ab_max_text}) for $R_{ab,\rm{max}}$ (black solid line), and Eq.\,(\ref{R_ab_min_text}) for $R_{ab,\rm{min}}$ (black dashed line).   

\subsection{Calculation results}
To further show the utility of the proposed readout scheme beyond the parameter values used in the experiment, we calculate $R$ and $F$ for a wide range of amplifier gain $G$ and added noise by the output chains as shown in Fig.\,\ref{RandFvsGandN}, where $N_{\rm{sys}} \equiv N_{\rm{sys},a}=N_{\rm{sys},b}$. In particular, we plot $R$ for the phase-preserving amplification case in panel \textbf{a} calculated using Eq.\,(\ref{R_a_uniform_r_text}), the maximized $R_{ab}$ for the combined mode in panel \textbf{b} calculated using Eq.\,(\ref{R_ab_max2_text}), and their ratio in panel \textbf{c}. As seen in panels \textbf{a} and \textbf{b}, we generally obtain higher power SNR for lower gains and higher noise levels when the two-mode squeezed state of the amplified readout signal is employed compared to reading out with a phase-preserving amplified mode. The same conclusion can be drawn from the $R$ ratio plotted in panel \textbf{c}, which is higher than unity for all practical values of $G>1$ dB and $N_{\rm{sys}}>0.5$, and can be as high as $1.66$ at $G=8$ dB and $N_{\rm{sys}}=50$.

Likewise, in panels \textbf{d} and \textbf{e} we showcase the general advantage in the assignment fidelity for the two-mode squeezed state readout ($F_{ab}$) over the phase-preserving amplification case ($F_a$). In panel \textbf{f}, we plot the fidelity difference ($\Delta F=F_{ab}-F_{a}$), which is positive for all practical values of $G>1$ dB and $N_{\rm{sys}}>0.5$. It is worth noting that although major improvements in the fidelity ($2-6\%$) are attained in the low to medium gain ($2-13$ dB) and medium to high added noise ($8-50$), meaningful fidelity improvements above $99.5 \%$ in the range $0.05-0.25 \%$ are still attainable at $G=20$ dB as seen in Fig.\,\ref{Fcrosssections}, which depicts cross sections of the calculated $F_{a}$ (black), $F_{ab}$ (blue) and $\Delta F$ (green) versus $N_{\rm{sys}}$ shown in Fig.\,\ref{RandFvsGandN} taken at fixed gains of $10$ dB (Fig.\,\ref{Fcrosssections} \textbf{a}) and $20$ dB (Fig.\,\ref{Fcrosssections} \textbf{b}). Such improvements in the readout fidelity above $99.5\%$ can be crucial in the pursuit of high-performance, fault-tolerant quantum computers that require as low readout errors as possible.   

\section{Discussion}

In this work, we enhance the readout fidelity of superconducting qubits, measured with phase-preserving quantum-limited amplifiers, by coherently combining the readout measurements taken simultaneously at the signal and idler frequencies such that we only amplify the combined mode quadrature carrying the qubit information. In other words, we boost the SNR of the qubit measurement setup by utilizing the quantum correlations between the signal and idler modes of the JM to enhance the coherent combined quadrature containing the qubit state information by a larger amount than the accompanying rise in the total noise of the output chain.

Here we also investigate the dependence of the measured average photon numbers of the signal, idler, and combined mode at the output of the JM as well as the SNR and fidelity metrics as a function of the average photon number at its input and the relative phase between the signal and idler modes. We further show that these results exhibit good agreement with the theory. 

The measurement scheme presented in this work exhibits three key advantages: (1) it does not require any changes to the quantum chip, the readout method, or the standard components used in the readout line, i.e., with the exception of being able to measure the idler mode of the amplifier on the same readout line or a separate one that has similar characteristics. (2) It exhibits enhancements in the resultant signal to noise ratio and readout fidelity compared to the standard phase-preserving amplification case for all practical values of amplifier gain and added noise by the output chain. (3) It is fully compatible with frequency-multiplexed amplification setups employed in large quantum processors.

It is important to emphasize that while in the experiment we use a particular nondegenerate three-wave mixing Josephson amplifier (JM) to demonstrate the enhancement in the qubit readout fidelity, the results are quite generic. They are not tied to any specific amplifier implementation, i.e., reflection-amplifiers like JMs, or directional amplifiers like TWPAs, nonlinear element, i.e., JJ \cite{DualPumpTWPA}, dc-SQUID \cite{JPASQUIDArrayJILA}, SNAIL \cite{SNAILresAmp}, arrays of rf-SQUIDs \cite{BroadbandSnakes}, JRM \cite{hybridLessJPC}, or mode of operation, i.e., three-wave mixing \cite{BroadbandCVKerrJMetamaterial} versus four-wave mixing \cite{ObsTwoModSqTWPA}. In particular, in the case of TWPAs where the signal and idler modes propagate on the same transmission line, it suffice to use only one output line (versus two used in this work) to measure the signal and idler modes using the room-temperature-stage electronics.

Furthermore, while we demonstrate here the benefits of using two-mode squeezed state of the measurement signal to improve the readout fidelity of superconducting qubits, the resultant enhancement in the signal to noise ratio of the measurement signal could potentially be applied in other quantum settings in which low-noise amplification is needed to detect  weak quantum signals in the microwave domain that are primarily encoded in one quadrature of the electromagnetic field, such as squeezed vacuum employed in the search of dark-matter axions \cite{
	AxionSearchNature,AxionSearchProposal}.
  
\section{Methods}

\subsection{Measurement parameters}

All measurement results included in the paper are taken using a heterodyne setup (see Fig. S2), in which output noise or readout signals exiting the fridge are downconverted, sampled, and digitized using a dual-channel Alazar card. Also, all microwave sources and measurement devices are phase locked to a $10$ MHz reference oscillator of a rubidium atomic clock.

In all 2D scatter data shown in the paper, we measure $N=10^4$ data points. In the experiment, we apply $1$ $\mu$s readout pulses and integration times $T_{\rm{int}}$ of the same duration. We also use a $10$ MHz frequency offset in the upconversion and downconversion of the input and output readout signals at the room-temperature stage (see Fig. S2). 

Other key measurement parameters of the qubit (transmon), readout, and JM device are listed in Table.\,\ref{ExpParams}.  

\begin{table}[tbh]
	\centering
	\begin{tabular}{||c |c ||} 
		\hline
		Parameter & Nominal value   \\
		\hline
		\hline
		$f_r$ (GHz) & $7.2284$      \\ 
		%\hline
		$\kappa/2\pi$ (MHz) & $1.91$    \\
		%\hline
		$\chi/2\pi$ (MHz) & $0.94$     \\
		$\theta$ & $53.7^{\circ}$ \\
		\hline
		$f_q$ (GHz) & $5.41227$   \\
		%\hline
		$\alpha_q$ (MHz) & $-332$   \\
		%\hline
		$T_1$ ($\mu s$) & $75$   \\
		%\hline
		$T_{\rm{2E}}$ ($\mu s$) & $52$   \\
		\hline
		$f_{a}$ (GHz) & $7.2284$  \\
		%\hline
		$f_{b}$ (GHz) & $9.7056$  \\
		%\hline
		$f_{p}$ (GHz) & $16.934$  \\
		%\hline
		$\gamma_{a}/2\pi$ (MHz) & $78$  \\
		%\hline
		$\gamma_{b}/2\pi$ (MHz) & $84$  \\
		%\hline
		$f_{a,\rm{max}}$ (GHz) & $7.36$  \\
		%\hline
		$f_{b,\rm{max}}$ (GHz) & $9.964$  \\
		\hline 
	\end{tabular}
	\caption{Readout, qubit (transmon), and JM parameters.  Readout resonator frequency ($f_r$), readout resonator bandwidth ($\kappa/2\pi$), state-dependent frequency shift $\chi/2\pi$, dispersive phase shift $\theta$, qubit frequency ($f_q$), qubit anharmonicity ($\alpha_q$), average $T_1$ and $T_{\rm{2E}}$ (echo) of the qubit, JM resonance frequencies on ports $a$ and $b$ at the flux working point ($f_a$ and $f_b$), pump frequency at the flux working point ($f_p$), JM linear bandwidths for mode $a$ and $b$ at the flux working point ($\gamma_{a}/2\pi$ and $\gamma_{b}/2\pi$), maximum resonance frequencies of the JM at zero applied flux ($f_{a,\rm{max}}$ and $f_{b,\rm{max}}$).}
	\label{ExpParams}
\end{table}

\subsection{Calibration of the measured quadratures}

To refer the output field quadratures measured using the room-temperature electronics back to the output of the JM, we employ the total gain of the output chain measured for the two different modes and lines (detailed in Sec. IV in the Supplementary Information).

The experimentally measured four dimensionless quadratures of the output fields corresponding to mode $a$ and $b$ at room temperature $I^{\prime}_a, Q^{\prime}_a, I^{\prime}_b, Q^{\prime}_b$ can be written as \cite{ObsTwoModSqTWPA}    

\begin{equation}
	I^{\prime}_{k}=\dfrac{A_{k}+A^{\dagger}_{k}}{2}, \quad Q^{\prime}_{k}=\dfrac{A_{k}-A^{\dagger}_{k}}{2i}, \label{RT_IQprime_quad}
\end{equation}

\noindent where $A_k$ and $A^{\dagger}_k$ are the annihilation and creation operators for mode $k$, respectively. They are obtained from the measured raw quadratures in \textit{Volts}, $I^{\rm{raw}}_{a}, Q^{\rm{raw}}_{a}, I^{\rm{raw}}_{b}, Q^{\rm{raw}}_{b}$ using the conversion relations

\begin{equation}
	I^{\prime}_{k}=\sqrt{\gamma_{k}}I^{\rm{raw}}_{k}, \quad Q^{\prime}_{k}=\sqrt{\gamma_{k}}Q^{\rm{raw}}_{k}, \label{RT_IQraw_quad}
\end{equation}

\noindent with the conversion factor $\gamma_{k}$ given by 

\begin{equation}
	\gamma_{k}=\dfrac{T_{\rm{int}}}{RE^{\rm{ph}}_{k}}, \label{convFactor}
\end{equation}

\noindent where $R=50$ Ohm, $T_{\rm{int}}$ is the integration time of the measurement, and $E^{\rm{ph}}_{k}=\hbar\omega_{k}$ is the photon energy of mode $k$.

Finally, we refer the measured quadratures $I_a, Q_a, I_b, Q_b$ back to the output of the device using the amplitude scaling 

\begin{equation}
	I_{k}=I^{\prime}_{k}/\sqrt{G_{\rm{sys},k}}, \quad Q_{k}=Q^{\prime}_{k}/\sqrt{G_{\rm{sys},k}}. \label{iq_quad}
\end{equation}

\textbf{Data availability} The data that support the findings of this study are available from the corresponding author upon request.

\section*{ACKNOWLEDGMENTS}
We are grateful to Andrew Eddins for valuable feedback and to David Lokken-Toyli and Corrado Mancini for discussions. We are also grateful to Nick Bronn and Vincent Arena for printed circuit board design, Sarunya Bangsaruntip for JM fabrication, and Daniela Bogorin for providing the cQED device. 

\textbf{Supplementary information}
Supplementary information is available.

\textbf{Author contributions} B.A. conceived and performed the experiments, developed the theory, and analyzed the data, W.S. developed the data-taking software, O.J. and J. R. made significant contributions to the experimental setup. B.A. wrote the paper with input from the other authors.

\textbf{Competing Interests} The authors declare no competing interests.

\end{document}

\title{Supplementary information for ``Enhancing qubit readout fidelity with two-mode squeezing of the coherent measurement signal"}
\author{Baleegh Abdo}
\author{William Shanks}
\author{Oblesh Jinka}
\author{J. R. Rozen}
\affiliation{IBM Quantum, IBM Research Center, Yorktown Heights, New York 10598, USA.}
\date{\today}

\maketitle
\newpage

\section{JM operation}
The output fields of the Josephson mixer (JM) are obtained by applying the unitary two-mode squeeze operator  $S=\exp{\left(re^{i\varphi_p}a^{\dagger}_{\rm{in}}b^{\dagger}_{\rm{in}}-re^{-i\varphi_p}a_{\rm{in}}b_{\rm{in}} \right) }$ to the input fields $a_{\rm{in}}$ and $b^{\dagger}_{\rm{in}}$ \cite{GenEntMwRadoverTL}, 

\begin{align}
	\begin{array}
		[c]{cc}%
		a_{\rm{out}}=S^{\dagger}a_{\rm{in}}S=\cosh\left(r \right) a_{\rm{in}}+e^{i\varphi_p}\sinh\left(r \right)b^{\dagger}_{\rm{in}}  ,\\
		b^{\dagger}_{\rm{out}}=S^{\dagger}b^{\dagger}_{\rm{in}}S=\cosh\left( r\right) b^{\dagger}_{\rm{in}}+e^{-i\varphi_p}\sinh\left( r\right) a_{\rm{in}},
	\end{array}
	\label{a_b_out_squeezer}
\end{align}

\noindent where $\varphi_p$ is the phase of the pump and the squeezing parameter $r\geq0$ relates to the device power gain by $G=\cosh^{2}(r)$.

Using Eq.\,(\ref{a_b_out_squeezer}), nonlocal two-mode squeezing effects emerge when evaluating the combinations of the output fields given by

\begin{equation}
	a_{\rm{out}}\pm e^{i\varphi_p}b^{\dagger}_{\rm{out}}=e^{\pm r}\left(a_{\rm{in}} \pm e^{i\varphi_p} b^{\dagger}_{\rm{in}} \right). \label{nonlocalsqueezing} 
\end{equation}

The field quadratures of the output bosonic modes $a_{\rm{out}}$ and $b_{\rm{out}}$ can be written in terms of the annihilation and creation operators,  

\begin{align}
	I_{a}=\dfrac{a_{\rm{out}}+a^{\dagger}_{\rm{out}}}{2}, 
	\quad Q_{a}=\dfrac{a_{\rm{out}}-a^{\dagger}_{\rm{out}}}{2i},  
	\nonumber \\ 
	I_{b}=\dfrac{b_{\rm{out}}+b^{\dagger}_{\rm{out}}}{2}, 
	\quad Q_{b}=\dfrac{b_{\rm{out}}-b^{\dagger}_{\rm{out}}}{2i}. 
	\label{IQxp_ab}
\end{align}

Similar expressions can be written for the input field quadratures $I_{a,\rm{in}}$, $Q_{a,\rm{in}}$, $I_{b,\rm{in}}$, and $Q_{b,\rm{in}}$ using the operators $a_{\rm{in}}$, $a^{\dagger}_{\rm{in}}$, $b_{\rm{in}}$, $b^{\dagger}_{\rm{in}}$.

In our experiment, $I_{a,\rm{in}}$, $Q_{a,\rm{in}}$ represent quadratures of the coherent state measuring the qubit, while $I_{b,\rm{in}}$, and $Q_{b,\rm{in}}$ correspond to quadratures of the input vacuum noise at port $b$. Thus, both input fields satisfy the commutation relations $[I_k,Q_k]=i/2$, where $k=a,b$. It follows that  $\left\langle \left( \Delta I_k \right) ^2\right\rangle=\left\langle \left( \Delta Q_k \right) ^2\right\rangle=1/4$ and satisfy the Heisenberg limit $\left\langle \left( \Delta I_k \right) ^2\right\rangle \left\langle \left( \Delta Q_k \right) ^2\right\rangle=1/16$, where  $\left\langle \left( \Delta I_k \right) ^2\right\rangle\equiv\left\langle I^{2}_k\right\rangle-\left\langle I_k\right\rangle^{2}$. 

Using Eqs.\,(\ref{a_b_out_squeezer}) and (\ref{IQxp_ab}), and assuming without loss of generality $\varphi_p=0$, we obtain 

\begin{align}
	I_{a}=\cosh\left( r\right)I_{a,\rm{in}}+\sinh\left( r\right) I_{b,\rm{in}}, 
	\label{I_a_vs_Iin_ab} \\ 
	Q_{a}=\cosh\left( r\right)Q_{a,\rm{in}}-\sinh\left( r\right) Q_{b,\rm{in}},  
	\label{I_b_vs_Iin_ab} \\ 
	I_{b}=\cosh\left( r\right)I_{b,\rm{in}}+\sinh\left( r\right) I_{a,\rm{in}}, 
	\label{Q_a_vs_Qin_ab} \\ 
	Q_{b}=\cosh\left( r\right)Q_{b,\rm{in}}-\sinh\left( r\right) Q_{a,\rm{in}}. 
	\label{Q_b_vs_Qin_ab}
\end{align}

By combining, for example, the output quadratures of both modes we obtain

\begin{align}
	I_{\rm{a}}+I_{\rm{b}}&=e^{r}\left( I_{a,\rm{in}}+I_{b,\rm{in}}\right), \label{Ia_p_Ib} \\
	I_{\rm{a}}-I_{\rm{b}}&=e^{-r}\left( I_{a,\rm{in}}-I_{b,\rm{in}}\right), \label{Ia_m_Ib} \\
	Q_{a}+Q_{b}&=e^{-r}\left( Q_{a,\rm{in}}+Q_{b,\rm{in}}\right), \label{Qa_p_Qb} \\
	Q_{a}-Q_{b}&=e^{r}\left( Q_{a,\rm{in}}-Q_{b,\rm{in}}\right),\label{Qa_m_Qb} 
\end{align}

\noindent which exhibit squeezing in the amplitude of the combined quadratures in Eq.\,(\ref{Ia_m_Ib}) and Eq.\,(\ref{Qa_p_Qb}) and antisqueezing along the orthogonal combined quadratures in Eq.\,(\ref{Ia_p_Ib}) and Eq.\,(\ref{Qa_m_Qb}). 

When considering each output field separately, we obtain an amplified thermal state with variance $\left\langle \left( \Delta a_{\rm{out}}\right) ^{2}\right\rangle =\left\langle \left( \Delta I_{a}\right) ^{2} \right\rangle+\left\langle \left( \Delta Q_{a}\right) ^{2}\right\rangle=\cosh\left( 2r\right)/2 $. However, when we calculate the variance for the combined quadratures $\left\langle \left(I_a-I_b\right) ^2\right\rangle=\left\langle \left(Q_a+Q_b\right) ^2\right\rangle =e^{-2r}/2$, we find that for any $r>0$, the noise variance is squeezed below the standard vacuum limit of $1/2$, while for $\left\langle \left(I_a+I_b\right) ^2\right\rangle=\left\langle \left(Q_a-Q_b\right) ^2\right\rangle =e^{2r}/2$ the noise variance is amplified above that limit. 

\section{Qubit readout signal at the JM output}

The average photon number at the output ports $a$ and $b$ of the JM, corresponding to the qubit states $i=g,e$, read

\begin{align}
	\bar{n}_{a,i}=\bar{I}^2_{a,i}+\bar{Q}^2_{a,i}, \label{n_a_i_bar} \\
	\bar{n}_{b,i}=\bar{I}^2_{b,i}+\bar{Q}^2_{b,i}. \label{n_b_i_bar}
\end{align}

For a dispersive readout with a phase shift of $\theta$ between the ground and excited states that are symmetrically centered around the positive $Q$ axis, the average $I$ and $Q$ quadratures of the input signal are given by $\bar{I}_{\rm{in}}=\sin\left( \theta/2\right)\sqrt{\bar{n}_{\rm{in}}}$ and $\bar{Q}_{\rm{in}}=\cos\left( \theta/2\right)\sqrt{\bar{n}_{\rm{in}}}$, respectively, where $\bar{n}_{\rm{in}}$ represents the average photon number at the input of the JM at port $a$, which approximately equals to the average photon number in the qubit measurement signal applied to the readout resonator. 

Using Eqs.\,(\ref{I_a_vs_Iin_ab})-(\ref{Q_b_vs_Qin_ab}) and  $\bar{I}_{b,\rm{in}}=\bar{Q}_{b,\rm{in}}=0$ for vacuum noise, we obtain for both qubit states

\begin{align}
	\bar{n}_{a}=\cosh^{2}(r)\bar{n}_{\rm{in}}, \label{n_a_bar} \\
	\bar{n}_{b}=\sinh^{2}(r)\bar{n}_{\rm{in}}, \label{n_b_bar}
\end{align}

\noindent which can also be written as

\begin{align}
	\bar{n}_{a}&=G\bar{n}_{\rm{in}}, \label{n_a_bar_G} \\
	\bar{n}_{b}&=\left(G-1\right) \bar{n}_{\rm{in}}. \label{n_b_bar_G_m1}
\end{align}

Furthermore, we can explicitly express the various quadrature averages for the $g$ state at the JM outputs as 

\begin{align}
	\bar{I}_{a,g}=\sqrt{G}\bar{I}_{\rm{in}},
	\quad 
	\bar{I}_{b,g}=-\sqrt{\left( G-1\right)}\bar{I}_{\rm{in}},  
	\nonumber \\ 
	\bar{Q}_{a,g}=\sqrt{G}\bar{Q}_{\rm{in}}, 
	\quad
	\bar{Q}_{b,g}=\sqrt{\left( G-1\right)}\bar{Q}_{\rm{in}},  
	\label{IQ_bar_ab_ge}
\end{align}

\noindent and those for the $e$ state as $\bar{I}_{a,e}=-\bar{I}_{a,g}$, $\bar{Q}_{a,e}=\bar{Q}_{a,g}$, $\bar{I}_{b,e}=-\bar{I}_{b,g}$, and $\bar{Q}_{b,e}=\bar{Q}_{b,g}$. 

When combining the respective quadratures $I_a, Q_a$ of the qubit measurements taken at output $a$ and the rotated quadratures $I^r_b, Q^r_b$ of the corresponding measurements taken at output $b$, we get 

\begin{align}
	I_{ab,i}\left(\phi \right)=I_{a,i}+I^r_{b,i}\left(\phi \right), \label{I_ab_i} \\
	Q_{ab,i}\left(\phi \right)=Q_{a,i}+Q^r_{b,i}\left(\phi \right), \label{Q_ab_i}
\end{align}

\noindent where $I^r_{b,i}, Q^r_{b,i}$ are obtained by rotating the measured quadratures $I_{b,i}, Q_{b,i}$ with a variable relative phase $\phi$   

\begin{align}
	I^{r}_{b,i}\left(\phi \right)=\cos\left( \phi\right)  I_{b,i}-\sin \left( \phi\right) Q_{b,i}, \label{I_r} \\
	Q^{r}_{b,i}\left(\phi \right)=\sin\left( \phi\right) Q_{a,i}+ \cos\left( \phi\right) Q_{b,i}. \label{Q_r}
\end{align}

Two special and distinct quadrature combinations are obtained for $\phi=\pi$ and $\phi=0$. In the first case ($\phi=\pi$), we obtain antisqueezing in $I_{ab}$ and squeezing in $Q_{ab}$ (denoted with superscripts $a$ and $s$, respectively)

\begin{align}
	I^{a}_{ab,i}=I_{a,i}-I_{b,i}, \label{I_out_antisqueezing} \\
	Q^{s}_{ab,i}=Q_{a,i}-Q_{b,i}, \label{Q_out_squeezing}
\end{align}

\noindent while in the second ($\phi=0$), we obtain squeezing in $I_{ab}$ and antisqueezing in $Q_{ab}$.

\begin{align}
	I^{s}_{ab,i}=I_{a,i}+I_{b,i}, \label{I_out_squeezing} \\
	Q^{a}_{ab,i}=Q_{a,i}+Q_{b,i}. \label{Q_out_antisqueezing}
\end{align}

To further simplify the notations for the combined readout signals, we define $I_{\Sigma}\equiv I^{a}_{ab}$, $I_{\Delta}\equiv I^{s}_{ab}$, $Q_{\Delta}\equiv Q^{s}_{ab}$, and $Q_{\Sigma}\equiv Q^{a}_{ab}$. 

Like the definitions of $\bar{n}_{a,i}$ (Eq.\,(\ref{n_a_i_bar})) and $\bar{n}_{b,i}$ (Eq.\,(\ref{n_b_i_bar})), we define fictitious average photon numbers for the combined quadratures given by 

\begin{align}
	\bar{n}_{ab,i}\left(\phi \right) =\bar{I}^2_{ab,i}\left(\phi \right)+\bar{Q}^2_{ab,i}\left(\phi \right). \label{n_ab_i_bar} 
\end{align}

In the special case of $\phi=\pi$ and independently of the qubit state $i$, we obtain using Eqs.\,(\ref{IQ_bar_ab_ge}), (\ref{I_out_antisqueezing}),(\ref{Q_out_squeezing}), and (\ref{n_ab_i_bar})

\begin{align}
	\bar{n}_{ab,\pi}=\left(\sqrt{G}+\sqrt{G-1} \right)^2 \bar{I}^2_{\rm{in}}+\left(\sqrt{G}-\sqrt{G-1} \right)^2 \bar{Q}^2_{\rm{in}}, \label{n_ab_bar_pi} 
\end{align}

\noindent which we can write in a compact form as 

\begin{align}
	\bar{n}_{ab,\pi}=e^{2r} \bar{I}^2_{\rm{in}}+e^{-2r} \bar{Q}^2_{\rm{in}}. \label{n_ab_bar_pi_r} 
\end{align}

As seen in Eq.\,(\ref{n_ab_bar_pi_r}) while the qubit-state dependent quadrature ($I$) is amplified, the $Q$ quadrature that carries very little information about the qubit state is squeezed. 

Likewise, in the case of $\phi=0$ and independently of the qubit state $i$ (to a large extent), we obtain using Eqs.\,(\ref{IQ_bar_ab_ge}), (\ref{I_out_squeezing}),(\ref{Q_out_antisqueezing}), and (\ref{n_ab_i_bar})

\begin{align}
	\bar{n}_{ab,0}=\left(\sqrt{G}-\sqrt{G-1} \right)^2 \bar{I}^2_{\rm{in}}+\left(\sqrt{G}+\sqrt{G-1} \right)^2 \bar{Q}^2_{\rm{in}}, \label{n_ab_bar_zero} 
\end{align}

\noindent which we can write in a compact form as 

\begin{align}
	\bar{n}_{ab,0}=e^{-2r} \bar{I}^2_{\rm{in}}+e^{2r} \bar{Q}^2_{\rm{in}}. \label{n_ab_bar_zero_r} 
\end{align}

Unlike Eq.\,(\ref{n_ab_bar_pi_r}), in Eq.\,(\ref{n_ab_bar_zero_r}) the qubit-state dependent quadrature is squeezed while the $Q$ quadrature is amplified.

\section{SNR metrics}

The power SNR of the qubit-state measurement is given by

\begin{equation}
	R\equiv\rm{SNR}^2=\left( \dfrac{\bar{I}_g-\bar{I}_e}{\dfrac{\sigma_g+\sigma_e}{2}}\right)^2, \label{R_def}
\end{equation}

\noindent where $\bar{I}$ ($\sigma$) is the mean (standard deviation) of the $I$ quadrature of the output coherent state carrying the qubit information. The subscripts $g$ and $e$ correspond to the qubit being in the ground and excited state, respectively.

Assuming for simplicity that $\bar{I}_g=-\bar{I}_e$ and $\sigma\equiv\sigma_g=\sigma_e$, we get

\begin{equation}
	R=\left( \dfrac{2\bar{I}}{\sigma}\right)^2, \label{R_uniform}
\end{equation}

\noindent where $\left| \bar{I}\right| \equiv\left|\bar{I}_g\right|=\left|\bar{I}_e\right|$.

To derive the $R$ figure of merit at output $a$ of the JM for which phase-preserving amplification takes place, we substitute $\bar{I}=\sqrt{G}\bar{I}_{\rm{in}}$, 
$\sigma=\sqrt{\left( 2G-1\right) \sigma^2_{0,I}+\sigma^2_{N,I}}$ in Eq.\,(\ref{R_uniform}), where $\sigma^2_{0,I}=1/4$ $\left( \sigma^2_{N,I}=N_{\rm{sys},a}/2\right) $ is the variance of the vacuum (output chain) noise along quadrature $I$, and $N_{\rm{sys},a}$ is the noise photons added by the output chain $a$ referred back to the HEMT input.    

This gives

\begin{equation}
	R_a=\dfrac{4G}{\left( 2G-1\right)/4+N_{\rm{sys},a}/2}\bar{I}^2_{\rm{in}}. \label{R_a_uniform}
\end{equation}

Equation (\ref{R_a_uniform}) can also be written in terms of $r$ as 

\begin{equation}
	R_a=\dfrac{\left( e^{2r}+e^{-2r}+2 \right) }{\left( e^{2r}+e^{-2r}\right) /8+N_{\rm{sys},a}/2}\bar{I}^2_{\rm{in}}. \label{R_a_uniform_r}
\end{equation}

In contrast, in the case of antisqueezing of the $I$ quadrature, we obtain

\begin{equation}
	R_{ab,\rm{max}}=\dfrac{2e^{2r} }{e^{2r}/4+N_{\rm{sys},e}/4}\bar{I}^2_{\rm{in}}, \label{R_ab_max}
\end{equation}

\noindent by substituting $\bar{I}=\bar{I}_{\Sigma}=e^r\bar{I}_{\rm{in}}$ and $\sigma=\sqrt{e^{2r}/2+N_{\rm{sys},e}/2}$ in Eq.\,(\ref{R_uniform}), where $N_{\rm{sys},e}$ is an effective added noise photons by the output chains for modes $a$ and $b$. Assuming uncorrelated Gaussian noise with zero mean for the two output chains, we obtain $N_{\rm{sys},e}=N_{\rm{sys},a}+N_{\rm{sys},b}$.

By comparing the $R$ figures obtained for the measured phase-preserving amplification of Eq.\,(\ref{R_a_uniform_r}) and the two-mode antisqueezing of Eq.\,(\ref{R_ab_max}), we find that for equal $\bar{I}_{\rm{in}}$ and $N_{\rm{sys},e}\leqslant 2N_{\rm{sys},a}$ (i.e., $N_{\rm{sys},b}\leqslant N_{\rm{sys},a}$), the latter exceeds the former in a wide range of practical gain and added noise parameters, with the exception of a very limited range when the output chain is nearly noiseless $N_{\rm{sys}}\approx 0$ or when the gain is very small $r\approx 0$. Moreover, while both figures become asymptotically equal in the finite added noise case and high-gain limit ($r\gg0$), $R_{ab,\rm{max}}$ remains higher than $R_{a}$ though yields diminishing returns as expected. 

Similarly, by substituting $\bar{I}=\bar{I}_{\Delta}=e^{-r}\bar{I}_{\rm{in}}$ and $\sigma=\sqrt{e^{-2r}/2+N_{\rm{sys},e}/2}$ in Eq.\,(\ref{R_uniform}), we obtain for the squeezing case  

\begin{equation}
	R_{ab,\rm{min}}=\dfrac{2e^{-2r} }{e^{-2r}/4+N_{\rm{sys},e}/4}\bar{I}^2_{\rm{in}}. \label{R_ab_min}
\end{equation}

\section{Calibration of the gain and noise of the output chain}

\begin{figure}
	[tb]
	\begin{center}
		\includegraphics[
		width=0.6\columnwidth 
		]%
		{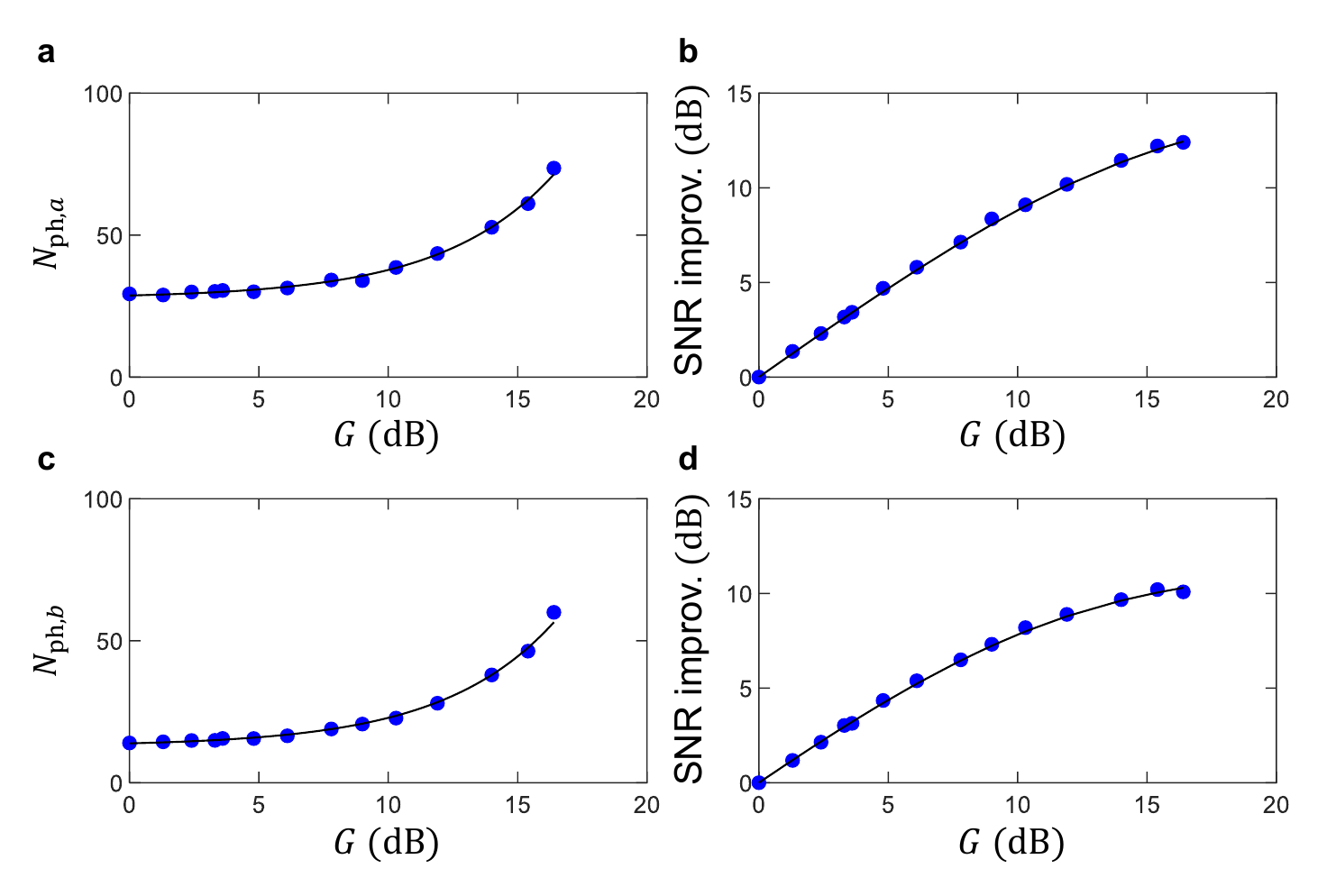}
		\caption{\textbf{Noise power measurements for calibrating the output lines for mode $a$ and $b$.} \textbf{a} (\textbf{c}) Noise equivalent photon number $N_{\rm{ph}}$ of mode $a$ ($b$) measured using the digitizer card versus JM gain $G$. The solid black curves are fits to the data which employ Eq.\,(\ref{G_Nsys1}). \textbf{b} (\textbf{d}) SNR improvement data for JM mode $a$ ($b$) versus $G$. The solid black curves are fits to the data which employ Eq.\,(\ref{SNR_improv}) and use the same $N_{\rm{sys}}$ parameters extracted in \textbf{a} (\textbf{b}). The extracted parameters from the fits are listed in Table \ref{GsysNphJM}.    
		}
		\label{OutCal}
	\end{center}
\end{figure}

\begin{figure*}
	[tb]
	\begin{center}
		\includegraphics[
		width=0.8\columnwidth 
		]%
		{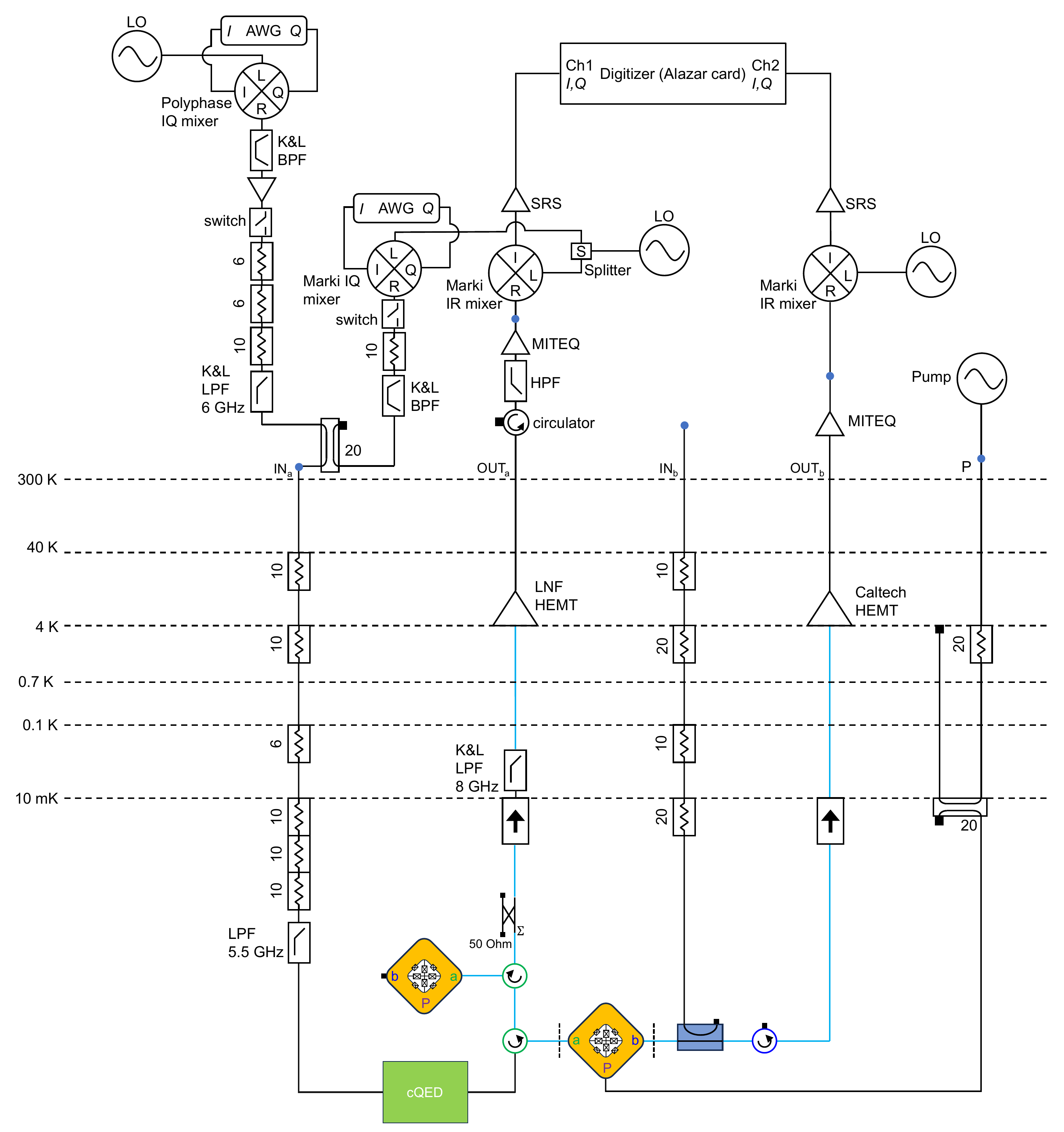}
		\caption{\textbf{Room temperature and dilution fridge measurement setup.} Qubit and readout pulses are combined at room temperature and sent to the cQED system through a single input line. The output of the cQED is coupled to port $a$ of the JM via a circulator. The amplified qubit readout signal is carried by two output lines $\rm{OUT}_a$ and $\rm{OUT}_b$ (coupled to ports $a$ and $b$ of the JM) and is simultaneously measured with a two-channel digitizer card at room temperature after being downconverted using IR mixers. The bandwidth of the green and blue circulators connected to port $a$ and $b$ are $4-8$ GHz, $8-12$ GHz, respectively. The cyan lines correspond to superconducting NbTi coax cables. The vertical dashed lines at the sides of the JM mark the reference planes for mode $a$ and $b$. The second (upper) JM was not used in the experiment and acts as a perfect mirror (pump off).     
		}
		\label{Setup}
	\end{center}
\end{figure*}

To calibrate the gain and noise of the output lines $a$ and $b$ of the JM, we employ the method used in Ref. \cite{EntSwap}.

The measured noise power of the output chain at room temperature is given by $P_{N,k}=\left( \sigma^{2}_{I,k}+\sigma^{2}_{Q,k}\right) /R$ \cite{ObsTwoModSqTWPA,BroadbandCVKerrJMetamaterial}, where the index $k=a,b$ corresponds to mode $a$ and $b$ of the JM, $\sigma^{2}_{I,k}$ and $\sigma^{2}_{Q,k}$ are the voltage noise variance of the $I$ and $Q$ quadratures of mode $k$, and $R=50$ Ohm is the input resistance of the measurement device.

At temperature $T$ at the bottom of the output line, the thermal photon population corresponding to angular frequency $\omega_k$ is given by the Bose-Einstein statistics  $1/2\coth\left(\hbar\omega_k/2k_BT \right)$. Since $T\cong10$ mK, and the mode frequencies are $\omega_a/2\pi=7.2284$ GHz and $\omega_b/2\pi=9.7056$ GHz, we obtain in the high-frequency/cold-temperature limit $\hbar \omega_k \gg K_BT$, $1/2\coth\left(\hbar\omega_k/2k_BT \right)\cong1/2$. 

Thus, the measured noise power of the output chain at room temperature as a function of $G$ reads

\begin{align}
	P_{N,k}\left(T\cong0, G \right)&=P_{0,k}\times N_{\rm{ph},k}\left(G\right) \nonumber
	\\
	& =P_{0,k} \times \left(G\dfrac{1}{2}+\left(G-1\right)\dfrac{1}{2} +N_{\rm{sys},k}\right), \label{G_Nsys1}
\end{align}

\noindent  where $P_{0,k}=G_{\rm{sys},k} \times BW  \times \hbar \omega_{k}$, $G_{\rm{sys},k}$ is the total gain/loss of the output chain for mode $k$, $BW=1/T_{\rm{int}}$ is the measurement bandwidth set by the integration time $T_{\rm{int}}$, and $N_{\rm{ph},k}\left(G\right)$ is the total output noise equivalent photon number for mode $k$ which varies with $G$.  

\begin{table}[tbh]
	\centering
	\begin{tabular}{|c|ccc|} 
		
		\hline
		Output & $G_{\rm{sys}}$ & $N_{\rm{sys}}$ & $T_{\rm{sys}}$ (K) \\ 
		\hline
		$\rm{OUT}_a$ &  $3.4\cdot10^6$ & 28.2 & 9.8 \\ 
		%\hline
		$\rm{OUT}_b$ &  $1.7\cdot10^7$ & 13.4 & 6.2  \\ 
		\hline 
	\end{tabular}
	\caption{Calibrated total gain and added noise of the output lines used in the experiment. $G_{\rm{sys}}$ and $N_{\rm{sys}}$ are extracted from fits to the data shown in Fig.\,\ref{OutCal}. The uncertainty in evaluating $G_{\rm{sys}}$ and $N_{\rm{sys}}$ is less than $\pm10\%$.} 
	\label{GsysNphJM}
\end{table} 

In the experiment, we measure the noise power using the same analog-to-digital converter (ADC) employed in the qubit readout. 

To find $G_{\rm{sys},k}$ and $N_{\rm{sys},k}$ for each output chain, we measure the output noise power for varying $G$ and extract their values by fitting the data using Eq.\,(\ref{G_Nsys1}). 

Another useful measure that we consider is the improvement in the signal-to-noise ratio of the output chain due to the JM given by $G/G_{N,k}$, where $G_{N,k}$ is the output noise ratio (or noise rise) corresponding to the JM being on versus off

\begin{align}
	G_{N,k}&=\dfrac{P_{N,k}\left(G \right) }{P_{N,k}\left(G=1 \right) } \nonumber \\
	&=\dfrac{T_{\rm{sys},k}+GT_{Q,k}+\left( G-1\right)T_{Q,k} }{T_{\rm{sys},k}+T_{Q,k}}, \label{G_N}
\end{align}

\noindent where $T_{Q,k}=\hbar \omega_k/2k_B$ is the equivalent temperature of the zero-point fluctuations at mode $k$, and $T_{\rm{sys},k}=N_{\rm{sys},k}\hbar \omega_{k}/k_B$ is the effective noise temperature of the output chain at mode $k$ in the absence of the JM. Note that the figure $G_{N,k}$, unlike $P_{N,k}$, is independent of the measurement bandwidth and system gain. It mainly depends on the total noise added by the system. 

Using Eq.\,(\ref{G_N}), the signal-to-noise ratio improvement due to the JM reads

\begin{equation}
	\dfrac{G}{G_{N,k}}=\dfrac{T_{sys,k}+T_{Q,k}}{T_{\rm{sys},k}/G+T_{Q,k}\left[1+\left(G-1\right)/G\right] }. \label{SNR_improv}
\end{equation}

While the relations (\ref{G_Nsys1}) and (\ref{SNR_improv}) are not independent, we, nevertheless, apply the extracted system noise from the fit of Eq.\,(\ref{G_Nsys1}) into Eq.\,(\ref{SNR_improv}) and confirm that the resulting curve fits the signal-to-noise ratio improvement data as well.

In Fig.\,\ref{OutCal}, we exhibit calibration measurements of output lines $\rm{OUT_a}$ and $\rm{OUT_b}$ used in the experiment. Figures \ref{OutCal} \textbf{a}, \textbf{b} (\textbf{c}, \textbf{d}) exhibit the measurement results for mode $a$ ($b$) taken for $\rm{OUT_a}$ ($\rm{OUT_b}$). The filled blue circles represent measured data, while the solid black curves are theory fits. In Figs.\,\ref{OutCal} \textbf{a}, \textbf{c} we plot the output noise equivalent photon number measurements as a function of $G$. Likewise, in Figs.\,\ref{OutCal} \textbf{b}, \textbf{d} we plot the corresponding SNR improvement measurements. The values of $G_{\rm{sys},k}$ and $N_{\rm{sys},k}$ employed in the fits are listed in Table.\,\ref{GsysNphJM}.

\section{Setup}
A schematic diagram of the setup used in the experiment is shown in Fig.\,\ref{Setup}. 

The setup includes two input and output lines denoted $\rm{IN_a}$, $\rm{IN_b}$ and $\rm{OUT_a}$, $\rm{OUT_b}$, respectively. It also includes a pump line denoted P, which is used for driving the bottom JM used in the experiment. The upper JM shown in Fig.\,\ref{Setup} is kept in the off state during the measurement, thus acting as a perfect mirror for the readout signals. 

Both input and output lines incorporate standard commercial microwave components that are commonly used in superconducting qubit experiments, such as cryogenic attenuators and low-noise HEMT amplifiers mounted inside the fridge, and IQ mixers and fast switches located outside the fridge.

The input line $\rm{IN_b}$ is mainly used for probing mode $b$ of the JM and verifying its resonance frequency and gain. In contrast, the input line $\rm{IN_a}$ is mainly used for applying qubit and readout pulses to the cQED. Both kinds of pulses, i.e., qubit and readout, are combined at the room-temperature stage and fed into $\rm{IN_a}$ using a directional coupler as depicted in the setup diagram. 

The pump line includes a wideband $20$ dB directional coupler at the base stage, which attenuates the pump tone that drives the JM. The attenuation is realized by directing a large portion of the pump power towards the $4$K stage where it gets dissipated in a $50$ Ohm termination.   

The blue dots on the input, output, and pump lines shown at the room-temperature stage, mark the locations at which the various lines get connected or disconnected to microwave generators and measuring devices, such as a vector network analyzer (VNA), spectrum analyzer (SA), and hetrodyne setup that is commonly used in dispersive readout of superconducting qubits. 

The black lines connecting the various components inside and outside the fridge correspond to normal-metal coax cables, whereas the cyan lines inside the fridge correspond to NbTi superconducting coax cables. The latter lines are primarily employed to minimize insertion losses at various sections of the output lines that precede the HEMTs which are known to degrade the SNR of output lines.

A small superconducting magnetic coil is attached to the Oxygen-Free-High-Conductivity copper package of the JM, which enables us to match the resonance frequencies of mode $a$ of the JM and the qubit readout resonator by flux tuning the JM. 

\section{Excess dephasing with gain}

\begin{figure}
	[tb]
	\begin{center}
		\includegraphics[
		width=0.6\columnwidth 
		]%
		{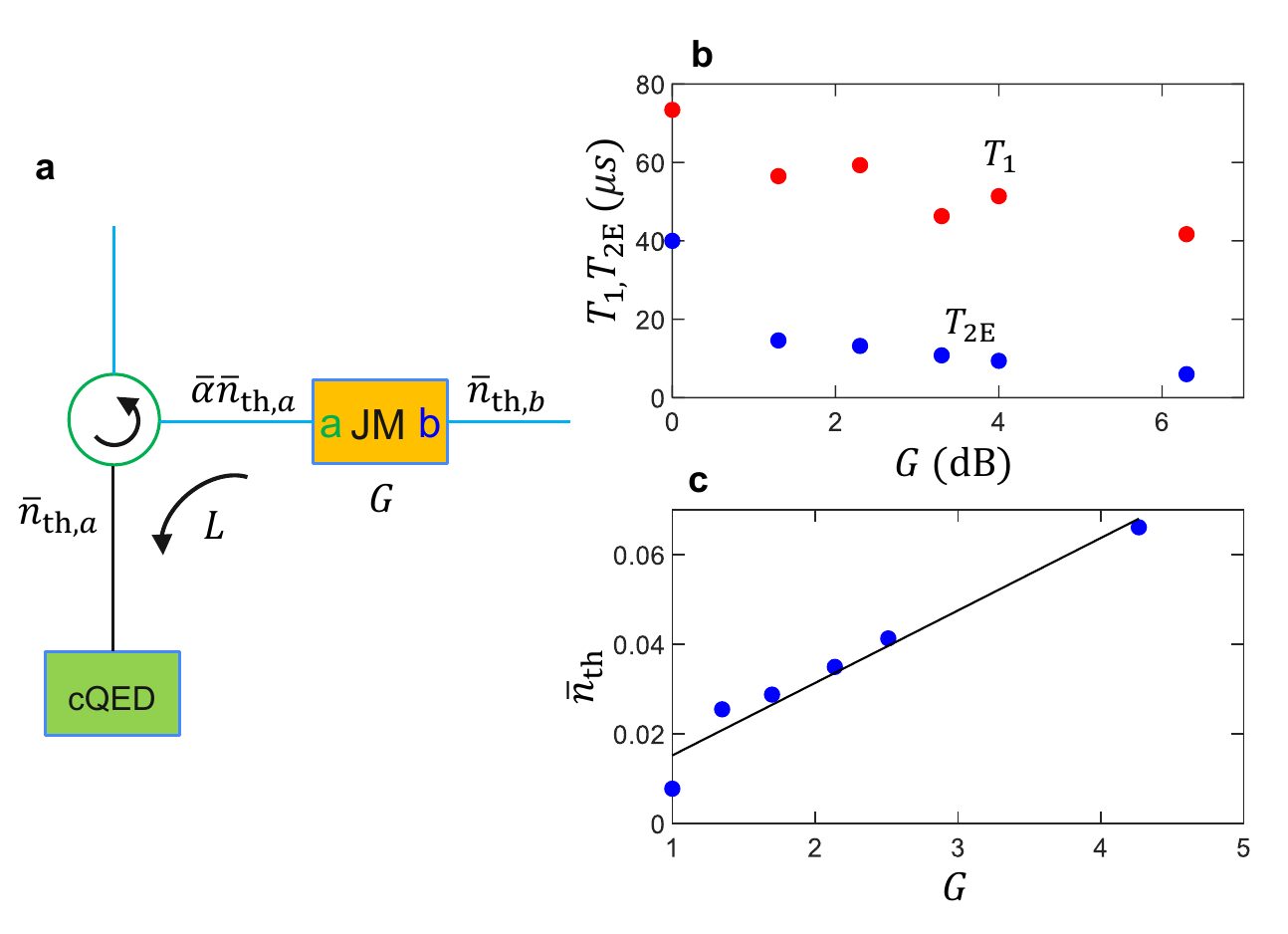}
		\caption{\textbf{Excess backaction on the qubit with JM gain.} \textbf{a} A block diagram showing the finite isolation between the JM and cQED. \textbf{b} Measurements of $T_1$ and $T_{\rm{2E}}$ of the qubit as a function of JM gain. \textbf{c} Average number of thermal photons in the readout resonator as a function of JM gain (blue discs). The linear fit (black) allows us to obtain using Eq.\,(\ref{L2}) an estimate for the parameter $L$, representing the isolation and insertion loss between the JM and cQED.       
		}
		\label{ThermalphVsG}
	\end{center}
\end{figure}

Due to the limited isolation between the JM and the cQED provided by one circulator as illustrated in Fig.\,\ref{ThermalphVsG} \textbf{a}, we observe a decrease in the qubit coherence times, i.e., $T_1$, $T_{\rm{2E}}$, with increased $G$ as shown in Fig.\,\ref{ThermalphVsG} \textbf{b}.  

From the measured $T_1$ and $T_{\rm{2E}}$ we can extract the dephasing time via the relation  $T_{\rm{\varphi}}^{-1}=T_{\rm{2E}}^{-1}-(2T_{\rm{1}})^{-1}$. Since thermal photon population in the readout resonator is the likely dominant dephasing mechanism in our dispersively coupled qubit-resonator system, the average number of thermal photons $\bar{n}_{\rm{th}}$ can be obtained from the dephasing rate $\Gamma_{\rm{\varphi}}\equiv T_{\rm{\varphi}}^{-1}$ in the limit of $\bar{n}_{\rm{th}}\ll 1$ using \cite{nbar}  

\begin{equation}
	\bar{n}_{\rm{th}}=\dfrac{\Gamma_{\rm{\varphi}}}{\Gamma_c},\label{n_th}\\
\end{equation}     

\noindent where $\Gamma_c=\kappa\chi^2/(\kappa^2+\chi^2)$, $\kappa$ is the total photon decay rate of the readout mode in the resonator, and $\chi$ is the qubit-state frequency shift of the readout mode. 

Using Eq.\,(\ref{n_th}) we plot in Fig.\,\ref{ThermalphVsG} \textbf{c}, the extracted $\bar{n}_{\rm{th}}$ versus $G$ (blue discs). By fitting the data to a line (black), we can estimate the total isolation and insertion loss between the JM and cQED, denoted $L$, using the relation 

\begin{equation}
	\bar{n}_{\rm{th}}=L\left[\bar{\alpha}\bar{n}_{\rm{th},a}+\bar{n}_{\rm{th},b}+1 \right]G-L\left[ \bar{n}_{\rm{th},b}+1\right]  +\bar{n}_{\rm{th},a},\label{L1}\\
\end{equation}

\noindent where $\bar{n}_{\rm{th},a}$ is the average number of thermal photons populating the cQED when $G=1$ (assuming $L\bar{\alpha}\ll1$) and $\bar{n}_{\rm{th},b}$ is the average number of thermal photons at the input of port $b$ of the JM. The unity terms in the first and second parentheses correspond to the effect of amplified vacuum noise at ports $a$ and $b$.

In the limit $\bar{n}_{\rm{th},a},\bar{n}_{\rm{th},b}\ll 1$, Eq.\,(\ref{L1}) reduces to

\begin{equation}
	\bar{n}_{\rm{th}}=LG+(\bar{n}_{\rm{th},a}-L),\label{L2}\\
\end{equation}

\noindent which gives $L\cong -18$ dB, in agreement with the general specifications of the circulator.